%% file: template.tex
\documentclass[journal]{vgtc}                
\ifpdf
  \pdfoutput=1\relax                   
  \usepackage{graphicx}                
  \DeclareGraphicsExtensions{.pdf,.png,.jpg,.jpeg} 
\else
  \usepackage{graphicx}                
  \DeclareGraphicsExtensions{.eps}     
\fi%

\graphicspath{{figures/}{pictures/}{images/}{./}} 

\usepackage{microtype}                 
\PassOptionsToPackage{warn}{textcomp}  
\usepackage{textcomp}                  
\usepackage{mathptmx}                  
\usepackage{times}                     
\usepackage{cite}                      
\usepackage{tabu}                      
\usepackage{booktabs}                  

\usepackage{hyperref}
\usepackage{adjustbox}
\usepackage{amssymb}
\usepackage{pifont}
\usepackage[tight,normalsize,sf,SF]{subfigure}
\usepackage{xcolor}
\usepackage{color, colortbl}

\usepackage{amssymb}
\usepackage{latexsym}

\usepackage{url}
\usepackage{xcolor}
\definecolor{newcolor}{rgb}{.8,.349,.1}

\usepackage{hyperref}

\onlineid{0}

\vgtccategory{Research}
\vgtcpapertype{please specify}

\PassOptionsToPackage{hyphens}{url}\usepackage{hyperref}
\usepackage{xcolor}
\definecolor{mycolor1}{HTML}{EDF8B1}
\definecolor{mycolor2}{HTML}{C7E9B4}
\definecolor{mycolor3}{HTML}{7FCDBB}
\definecolor{mycolor4}{HTML}{41B6C4}
\definecolor{mycolor5}{HTML}{1D91C0}
\definecolor{mycolor6}{HTML}{225EA8}

\title{Enhancing Vascular Analysis with Distance Visualizations: \\ An Overview and Implementation}

\author{Jan Hombeck$^{a*}$, Monique Meuschke$^b$, Simon Lieb$^a$, Nils Lichtenberg$^c$, Felix Fleisch$^a$, Maximilian Enderling$^a$, \\ Rabi Datta$^d$, Michael Krone$^c$, Christian  Hansen$^b$, Bernhard  Preim$^b$ and Kai Lawonn$^a$ \\ \\
\fontsize{07}{06}\selectfont
$^a$ University of Jena \\
\fontsize{07}{06}\selectfont
$^b$ University of Magdeburg \\
\fontsize{07}{06}\selectfont
$^c$ University of Tübingen \\
\fontsize{07}{06}\selectfont
$^d$ University Hospital Cologne 
}
\authorfooter{
\item $^*$jan.hombeck@uni-jena.de

}

\shortauthortitle{Hombeck \MakeLowercase{\textit{et al.}}: Enhancing Vascular Analysis with Distance Visualizations}

\abstract{In recent years, the use of expressive surface visualizations in the representation of vascular structures has gained significant attention. These visualizations provide a comprehensive understanding of complex anatomical structures and are crucial for treatment planning and medical education. However, to aid decision-making, physicians require visualizations that accurately depict anatomical structures and their spatial relationships in a clear and well-perceivable manner. This work extends a previous paper and presents a thorough examination of common techniques for encoding distance information of 3D vessel surfaces and provides an implementation of these visualizations. A Unity environment and detailed implementation instructions for sixteen different visualizations are provided. These visualizations can be classified into four categories: fundamental, surface-based, auxiliary and illustrative. Furthermore this extension includes tools to generate endpoint location for vascular models. Overall this framework serves as a valuable resource for researchers in the field of vascular surface visualization by reducing the barrier to entry and promoting further research in this area. By providing an implementation of various visualizations, this paper aims to aid in the development of accurate and effective visual representations of vascular structures to assist in treatment planning and medical education.

}


\vgtcinsertpkg

\begin{document}


\maketitle

\input{chapter/01_introduction.tex}
\input{chapter/02_perception.tex}
\input{chapter/03_medicalbackground.tex}

\input{chapter/04_implementationsetup.tex}
\input{chapter/05_paperselection.tex}

\input{chapter/06_shading_fund.tex}

\input{chapter/07_surface_techniques.tex}
\input{chapter/08_auxiliary_tools.tex}

\input{chapter/09_illustrative.tex}
\input{chapter/10a_LIC.tex}

\input{chapter/10b_Enpoints.tex}

\input{chapter/11_discussion.tex}
\input{chapter/12_conclusion.tex}

\acknowledgments{
This work was partly funded  from the
German Research Foundation (DFG) (Project No.: 290703152). 
}

\bibliographystyle{abbrv-doi}

\bibliography{template}
\end{document}

%% file: chapter/01_introduction.tex
\section{Introduction}
Expressive visualizations of vascular structures are essential for treatment planning and training in interventional radiology and surgery. An expressive visualization of a vascular tree conveys its branching pattern well and makes anatomical variants obvious. Vascular structures exhibit special properties in comparison with other anatomical structures, such as organs, that motivate special visualization techniques beyond the common surface and volume rendering techniques. As an example, the shrinking effect of surface smoothing may lead to disconnected vascular trees. 
Specialized vessel visualization techniques may incorporate model assumptions such as the connectedness of a vascular tree, circular cross sections, and a continuous change of the local vessel diameter. Implicit surface visualizations are superior to explicit surfaces in terms of smoothness \cite{Preim2008}.
 
Research on vessel visualization initially was focused on technical aspects. Visualization techniques were developed that aimed at an accurate representation of vascular surfaces, at high smoothness (measured as curvature), and at a high performance. Although these aspects are still important, the question arises how different vessel visualization techniques or different parameterizations of the same technique affect the \emph{perception} of vascular structures. The whole motivation for 3D visualizations is to provide a good depth and shape perception. Starting with the seminal work of Ropinski~\cite{Ropinski2006} there was a series of publications on psycho-physical studies to investigate these perceptual aspects of vessel visualizations. While Ropinski's major idea was to employ color for depth encoding, other visualization techniques, such as textures and hatching, were employed as well.
 
The importance of perception-based vessel visualization further increased in recent years. Vessel visualizations were developed that convey not only the vascular surface but also associated scalar fields, such as wall thickness or wall shear stress derived from a flow simulation. Finally, due to the growing use of VR in surgical and interventional training advanced vessel visualizations are also explored in immersive settings. This further increases the demands to carry out psycho-physical experiments to understand the perceptual and cognitive consequences of using certain vessel visualization techniques.
 
The conduction of psycho-physical experiments involves various challenges, e.g. the careful selection of appropriate stimuli. Due to the importance of this process, research was carried out to support this process with automatically generated suggestions. The \textsc{EvalViz} \cite{Meuschke:2019:Eval} framework is the most notable example of this research.

We provide an extension of a previous paper \cite{hombeck2022distance} that included an overview of the most commonly used vessel visualizations for egocentric and exocentric distance estimation and includes an implementation setup for each visualization (see here  \footnote{\href{https://github.com/jhombeck/DepthVis}{\color{blue}Click Here: Unity Project}}). As an extension to the original paper, we introduce several visualization types that play a critical role in analyzing vascular data. These include visualizations for scalar and vector fields, as well as a hatching technique, which can now be directly applied from within the framework.

To further enhance the utility of the framework, we have also included endpoint detection for vascular structures. This is crucial in various clinical applications, such as diagnosis and surgical planning, as it enables the use of previously established visualizations without manual selection of glyph placement. 

In summary, this extension aims to improve the quality and usability of previous visualizations while incorporating additional crucial visualizations necessary for effective vascular analysis and promoting new research within this area, as these visualizations are now more easily accessible and bundled in one place. In addition, we provide an overview of the visual cues used in each of these visualizations. This way, a more informed selection of visualizations can be made. In general, this work can be seen as a starting point to promote further and more in-depth research in the field of vessel visualization. 
 

%% file: chapter/02_perception.tex
\section{Visual Perception}

The human visual perception is based on light-sensitive cells (rods and cones) in the retina \cite{Gibson1950}. Due to the high density of these cells in the fovea, we can perceive a high level of detail in a small region of our visual field. Visual perception occurs in two stages: the \emph{pre-attentive} stage, where we are able to detect objects that are strongly different from others, and the \emph{attentive} stage, where we search for certain objects and their relations based on our current goals. This conscious process requires considerably more time. As an example, a vessel segment that is emphasized with a unique color would be perceived pre-attentively, whereas the search for a trifurcation in a vascular tree requires attentive processing. Visual perception has many aspects, such as color, contrast, and motion perception. For our purpose of studying the perceptual effects of vessel visualizations, shape, and depth perception are particularly important and will be discussed in the following.

Although the light-sensitive cells are arranged in a planar configuration, we perceive the world as three-dimensional. Various depth cues, including  shadows, shading, perspective 
projection, and partial occlusion~\cite{Preim2016}
, enable us to infer the spatial configuration, the depth relation, and the object shapes with its bumps and bulges.
In addition to these \emph{monoscopic} depth cues, there is also \emph{stereoscopic vision} that provides further depth cues based on the combination of the two slightly different images perceived by both eyes. Stereoscopic vision only works in a certain depth range. However, when viewing visualizations on a desktop we are located in this range. Thus, stereoscopy is essential for medical visualization.

Virtual reality, where users typically wear VR glasses increases 
\emph{immersion}, i.e., users are completely surrounded by a virtual environment (VE) instead of just having a monitor as a window to a VE. This is beneficial for surgical training, where users carry out manipulations of the patient's anatomy. Currently, widespread VR glasses are still not optimal with respect to human perception. Their field of view (FOV) is smaller than the FOV that our visual system enables. Also, the spatial resolution of the VR glasses is not as high as our eyes would be able to perceive. Despite these limitations, the immersion and the motivating effect of VR already enables strong learning experiences \cite{Chheang:2019}.








\section{Previous and Related Work}
In this paper, we focus on 3D surface visualizations, especially of vascular structures. 
Therefore, we do not discuss map-based visualizations (see the recent survey of 
Eulzer et al.~\cite{Eulzer:2022}) and volume renderings of vascular structures 
(see the survey from Kubisch et al.~\cite{Kubisch:2012}). 
This focus is motivated by the wide use of surface visualizations in VR.
Furthermore, we focus on techniques developed for the perception of distances. 

\subsection{Surface Visualization of Vascular Structures} 
Surface visualizations are based on an explicit segmentation of vascular structures, 
often followed by the determination of the vessel centerline, the local vessel 
diameter, and the branching graph~\cite{Selle:2002}. 
This information may be employed to fit graphics primitives, such as cylinders 
or truncated cones, to the vessel centerline and ensure a continuous transition 
at strongly curved regions and at branchings. 
Pioneering work in this direction was carried out by Ehricke et al.~\cite{Ehricke:1994} 
and Puig et al.~\cite{Puig:1997}. 
As an alternative to explicit visualizations, implicit vessel visualizations were 
created. 
Care is necessary to avoid or reduce the typical ``unwanted'' effect of implicit 
surfaces, such as bulging and strong blending~\cite{Oeltze:2004,Schumann:2007}. 
More recent work aims at watertight meshes~\cite{Kretschmer:2013} and at optimizing 
the resulting geometry with a curvature-based polygon construction~\cite{Wu:2010}.

While a lot of effort was spent on optimizing the geometry, the performance, and 
accuracy of vessel surface visualizations, the actual rendering was largely neglected. 
Only recently, Ostendorf et al.~\cite{Ostendorf:2021} and Hombeck et al.~\cite{hombeck2022evaluating,hombeck2019evaluation,hombeck2022heads} studied the effect of different shading styles. 

\subsection{Distance Perception}
The correct estimation of distances plays an important role in treatment planning, 
e.g., to find well-suited access without violating risk structures.  
Here, two types of distances are distinguished: egocentric and exocentric distances. 
Egocentric distances describe distances between the viewer and their environment~\cite{Renner2013, ElJamiy2019a}, which is also known as depth perception. 
In contrast, exocentric distances relate to distances between two objects other than the viewer. 
In the following, we summarize visual cues to support distance perception. 

\paragraph*{Perception of egocentric distances}
Based on the visual information captured by the eyes, humans can judge distances 
to objects~\cite{Abhari2015visual}. 
The visual system must interpret various monocular and binocular cues to enable 
spatial and 3D depth perception, e.g., to judge the distance, depth, and shape of 
objects. 
Reichelt et al.~\cite{Reichelt2010}  distinguished two types of depth information 
of the visual system: oculomotor and visual depth information.
Oculomotor depth information relies on the capabilities of the eye, such as changing the angle of rotation of the eyes, the shape of the lens, and the tension of the eye muscles 
to perceive depth information. 
In this work, however, we focus on visual depth information in the form of 
different visualization techniques applied to 3D surfaces. 

There are \textit{monoscopic} and \textit{stereoscopic} depth cues. 
In the case of monoscopic cues, an open eye is sufficient to perceive the scene. 
They  can  be  divided into \textit{static} and \textit{motion-based} cues. 
The most important static depth cues include shadows, shading, perspective 
projection, and partial occlusion~\cite{Preim2016}.
Motion parallax is another essential depth cue: When we interact with a 3D model or observe how it is continuously rotated in an animation, we may infer depth relations. As an example, a static maximum intensity projection (MIP) does not provide any depth cues. However, in an animation, the vascular anatomy---highlighted by MIP---can be assessed well. 

Stereoscopic cues are a natural way to perceive egocentric distances with both eyes.
However, in case of printouts are needed or dynamic visualizations would require a high degree of interaction (e.g., during an operation), static images are 
irreplaceable. 
In these cases, additional depth information is essential.
Further subcategories of depth cues include motion-, surface- and 
illumination-based cues.  
Common techniques for these are color scales~\cite{Steenblik1987, Ropinski2006, kreiser2018void}, 
glyphs~\cite{LichtenbergHL17} or illustrative line drawings~\cite{Lawonn_2015_Feature,Lawonn:2018:SSI}. 

\paragraph*{Perception of exocentric distances.}
While several works address the perception-based encoding of depth 
information, there is little work on the visual representation of 
exocentric distances.  
There are works that research how to encode exocentric distances between a virtual 
needle and surroundings risk structures, such as vessels. 
Different visualization techniques, such as color-encoding~\cite{hombeck2022evaluating} 
or illustrative renderings~\cite{Lawonn_2015_MICCAI}, are used to explicitly encode 
the spatial distance between the needle and a second object. 
Similar techniques can also be used to encode exocentric distances between two 
3D surface models extracted from clinical image data, such as a vessel branch and 
a tumor. 
Wartenberg and Wiborg \cite{Wartenberg2003} evaluated the perception of exocentric 
distances in a desktop and an immersive cave environment. 
Here, the advantages of the cave setup were revealed. However, distances were rather 
overestimated.
In contrast to explicitly encoding exocentric distances, shape cues can be employed 
to improve the shape perception of objects that in turn support the perception of 
exocentric distances. 
Commonly used shading techniques to improve shape perception are Phong and Toon 
shading~\cite{Preim2016}.


%% file: chapter/03_medicalbackground.tex
\begin{figure*}
\centering
\begin{minipage}{.33\textwidth}
  \centering
  \includegraphics[width=\linewidth]{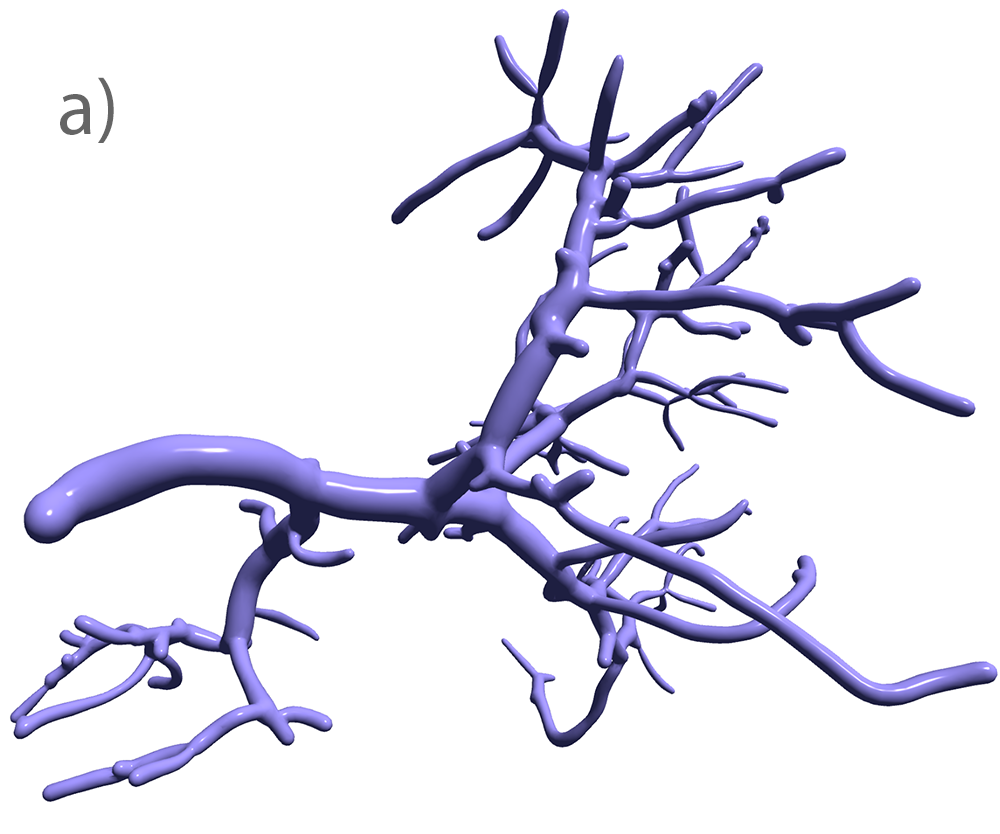}
  \label{fig:phong1}
\end{minipage}%
\begin{minipage}{.33\textwidth}
  \centering
  \includegraphics[width=\linewidth]{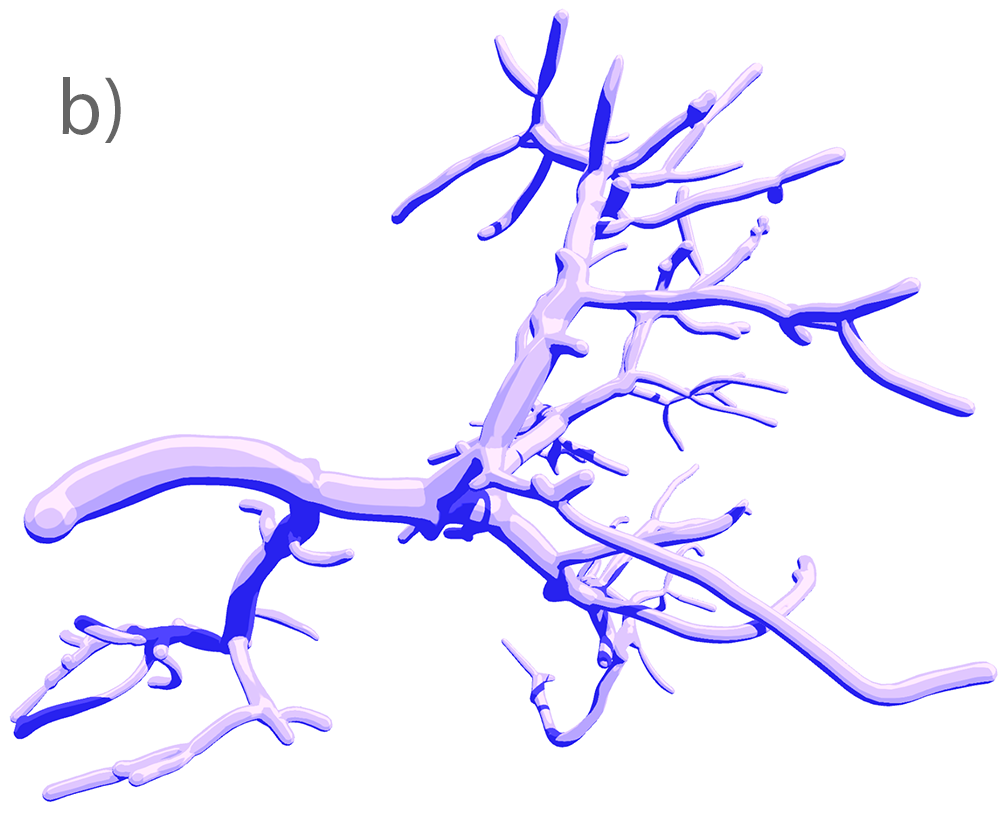}
  \label{fig:toon}
\end{minipage}
\begin{minipage}{.33\textwidth}
  \centering
  \includegraphics[width=\linewidth]{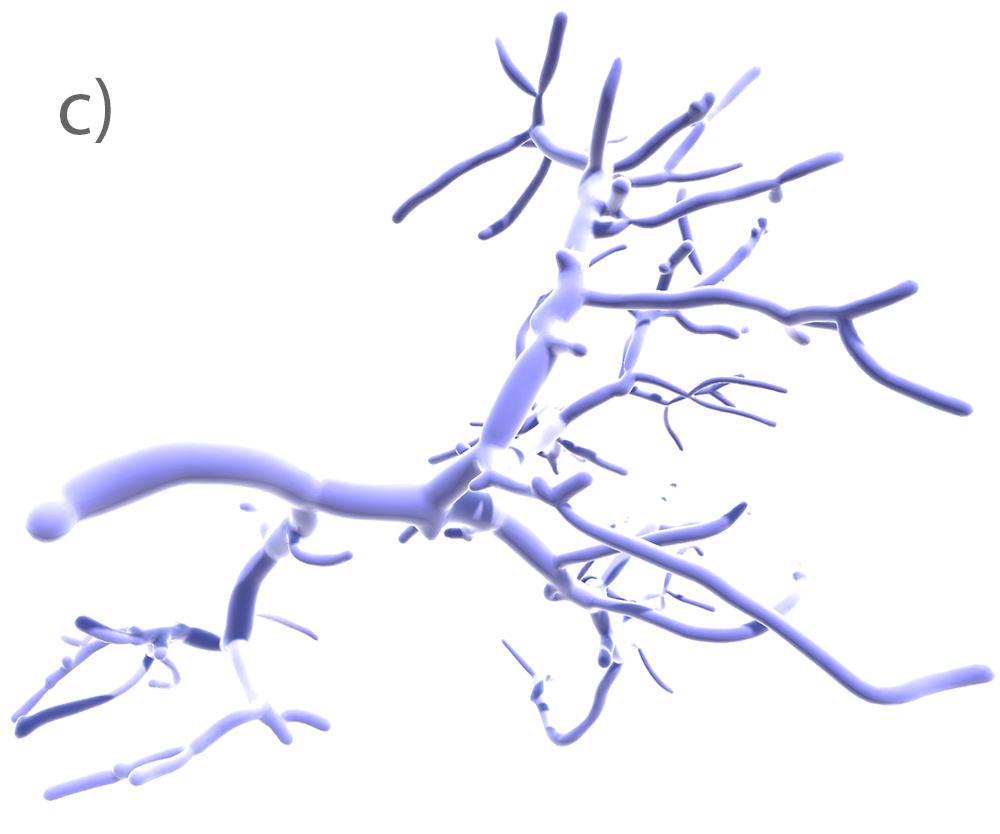}
  \label{fig:fresnel}
\end{minipage}
\caption{The fundamental shading techniques (a) Phong, (b) Toon, and (c) Fresnel effect applied on a  vascular structure.}
\label{fig:fundamentals}
\end{figure*}
\section{Medical Background }
Perception-based visualization plays a major role in the context of medical data. 
In clinical routine, 3D medical visualizations fulfill several tasks. 
They provide an overview of complex anatomical structures or fractures. 
Furthermore, they support treatment planning, e.g., the decision whether a tumor 
can be removed, the extent of the operation, and the possible access paths.
As an example, the surgical removal of a liver tumor must be carefully planned in  order to safeguard 
good surgical outcome ~\cite{Hansen2009,Hansen2014}. 
For this purpose, it is essential to understand the morphology of the organ and 
internal vasculature to determine the resection volume while preserving as much 
of the vasculature as possible.
Therefore, spatial relationships and the distances between structures should be 
made apparent. 

\paragraph*{Data Acquisition}
To support decision-making when dealing with complex anatomy and pathology, 
high-quality 3D models of organs and vessels can be reconstructed from medical 
image data. 
In this work, we use 
oncological liver surgery as an 
example to provide the implemented depth encodings. 
We used two liver data sets provided by our clinical partners. 
Each data set comprises a 3D surface mesh of the liver, the vessel tree, and a 
liver tumor.
The 3D surface models of these three structures comprise approx. 16,000, 
30,000, and 5,000 vertices, respectively. 
All data sets were segmented from a contrast-enhanced computed tomography scan 
by employing threshold-based segmentation. 
From the binary segmentation masks, 3D models of the corresponding structures are 
reconstructed using Marching Cubes. 

\subsection{Endpoint Selection}
Endpoint detection for 3D vascular objects, such as liver vessel trees, has been a topic of research in the medical imaging field as well as the 3D reconstruction area. This is due to the importance of accurately identifying the endpoints in such structures for various clinical applications, including diagnosis and surgical planning. 

A number of different approaches have been proposed for endpoint detection in 3D vascular objects. One popular method involves the use of image segmentation algorithms\cite{ardizzone2008blood,olsen2011convolution}, which separate the vessel tree from the surrounding tissue. This enables the identification of the endpoint as the tips of the tree-like structure. Another approach is to use the vessel tree as a graph and extract endpoint information based on the graph properties \cite{roy2019graph,dashtbozorg2013automatic}. This method is effective in cases where the vessel tree has a complex branching pattern.
When the 3D shape of the object is known and accessible, the endpoints of the mesh can be identified through the process of thinning and skeletonizing the mesh \cite{lichtenberg2020parameterization,guo2020novel}.

%% file: chapter/04_implementationsetup.tex
\section{Implementation Setup}
For many publications, it is difficult to replicate the presented work due to the lack of crucial implementation details. In this work, we provide a unified framework for selecting visualizations. This framework is intended to lower the entrance hurdle for performing experimental visualizations. This can increase their overall usability even for smaller projects. 
%
Within the collaboration with our clinical partners, we discovered that numerous research projects focusing on visualization, particularly in virtual reality, employ 3D engines, such as Unity, to create prototypes for specific scenarios.
With this and the cross-platform accessibility in mind, we chose Unity as our development platform. 
Unity is a versatile game engine that offers fundamental computer graphics commands that are cross-platform compatible and feature WebGL functionality to facilitate the creation of web-based 3D graphics. This way, even users without in-depth computer graphics knowledge can work with the visualizations presented. Unity not only supports Mac OS, Linux, and Windows but also provides a virtual and augmented reality interface that supports almost all devices available. 

As we intend to expand this framework in the future we have decided to focus on the long-term support version of Unity 2021.3 LTS. 
Once the correct Unity version is installed, our framework needs to be downloaded (see here \footnote{\href{https://github.com/jhombeck/DepthVis}{\color{blue}Click Here: Unity Project}}). 
Each visualization has its own Unity scene that can be loaded separately. This scene contains only the code needed for the particular visualization. Simple visualization requires only a material and a shader file in Unity. The shader is assigned to the material, which again is assigned to the desired 3D mesh. If the visualization has additional properties, such as color, brightness, and size, these properties can be changed within the material. These parameters can also be changed at runtime, making it easier to customize the visualization. More complex visualizations may also include a C\# script to execute additional functionality and/or pass uniform variables to the shader. Because some visualizations require information provided by scripts, they may appear incorrect in the editor but operate properly in the Unity game window. Each file for the visualization is open source and can be customized.

%% file: chapter/05_paperselection.tex

%% file: chapter/06_shading_fund.tex
\begin{figure*}
\centering
\begin{minipage}{.245\textwidth}
  \centering
  \includegraphics[width=\linewidth]{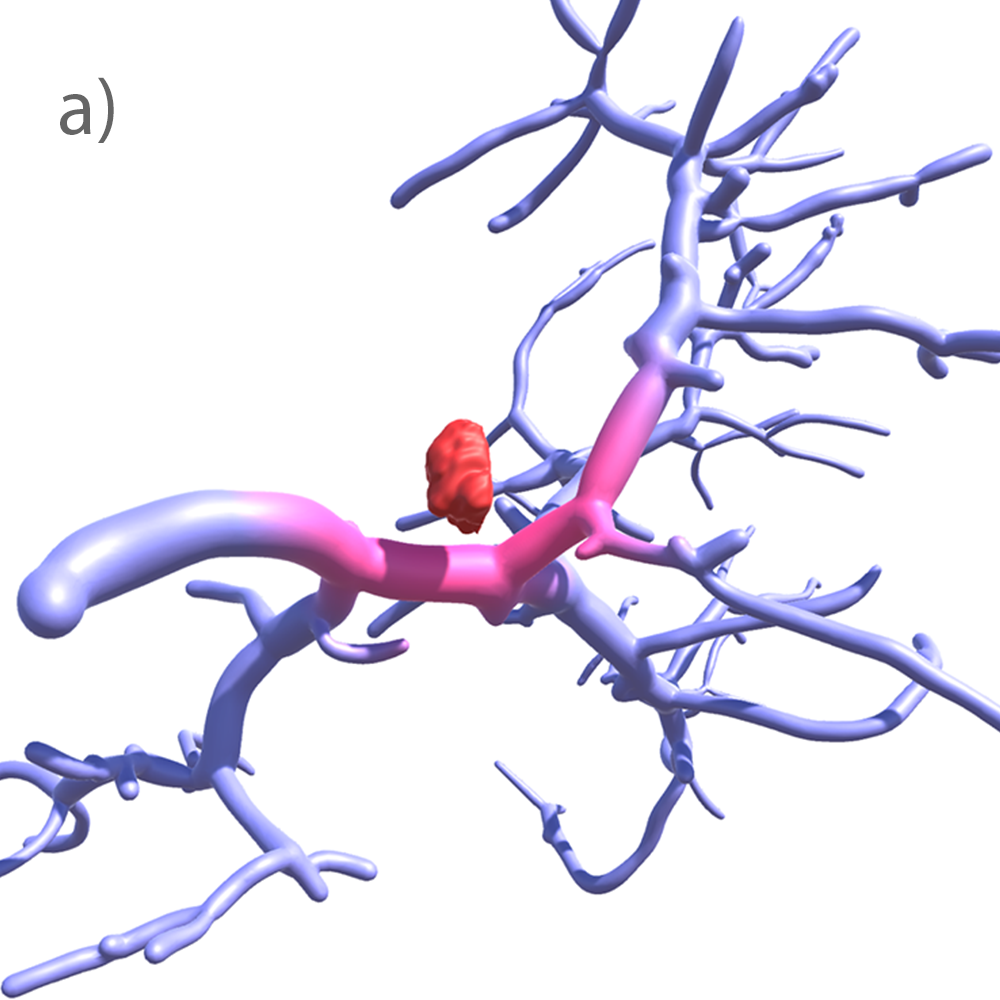}
  \label{fig:heatnap}
\end{minipage}%
\begin{minipage}{.245\textwidth}
  \centering
  \includegraphics[width=\linewidth]{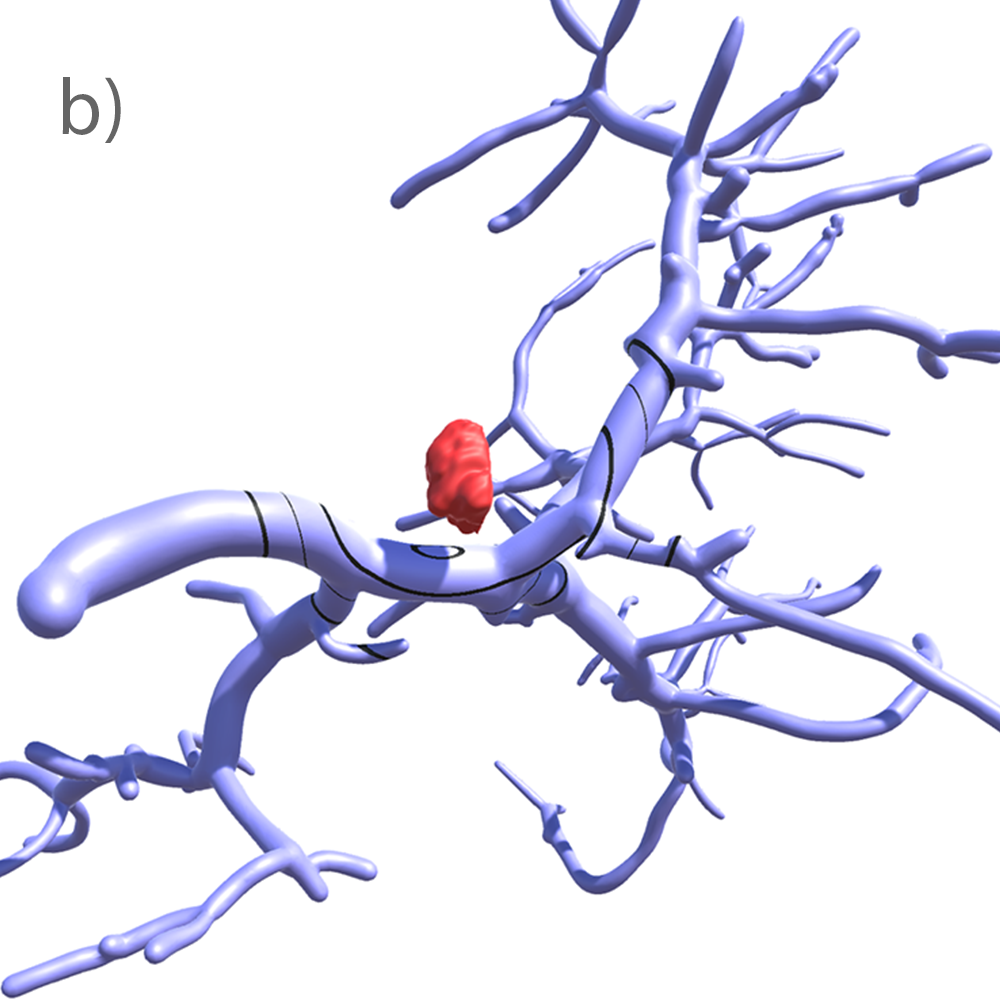}
  \label{fig:isolines}
\end{minipage}
\begin{minipage}{.245\textwidth}
  \centering
  \includegraphics[width=\linewidth]{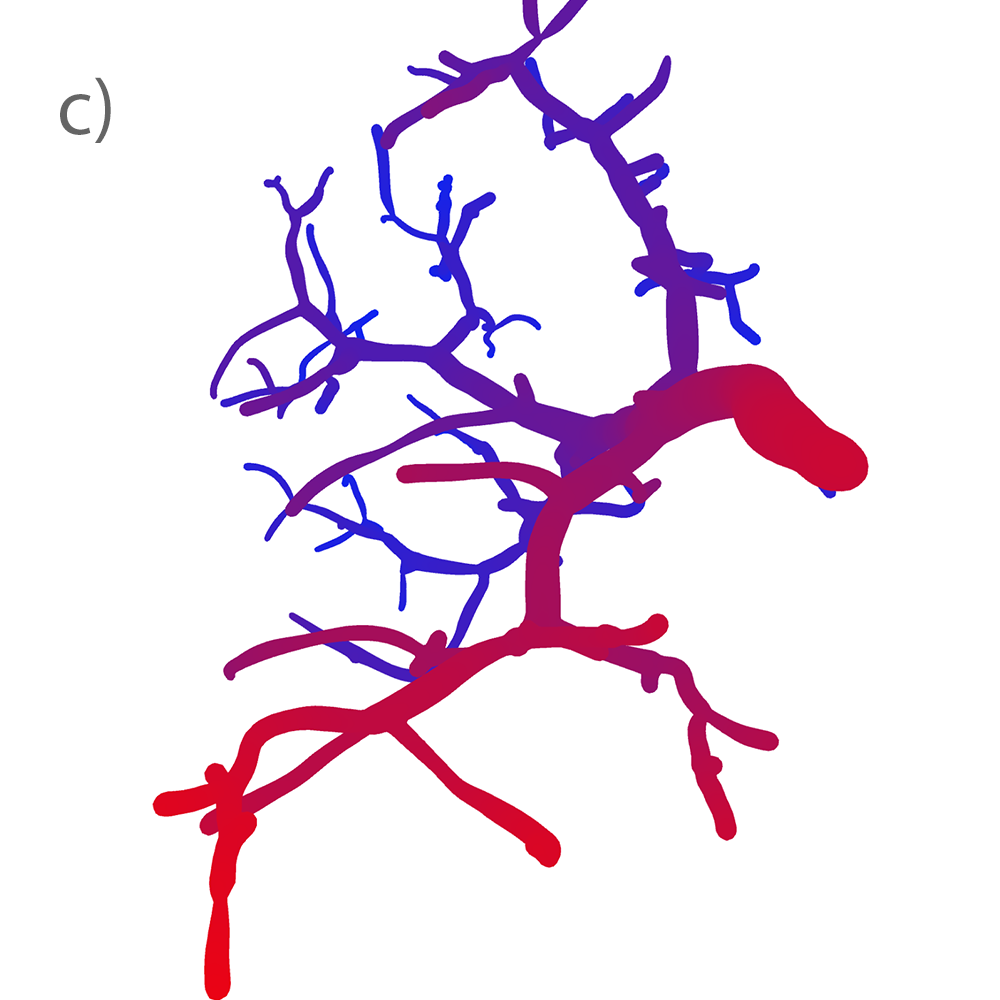}
  \label{fig:pcd}
\end{minipage}
\begin{minipage}{.245\textwidth}
  \centering
  \includegraphics[width=\linewidth]{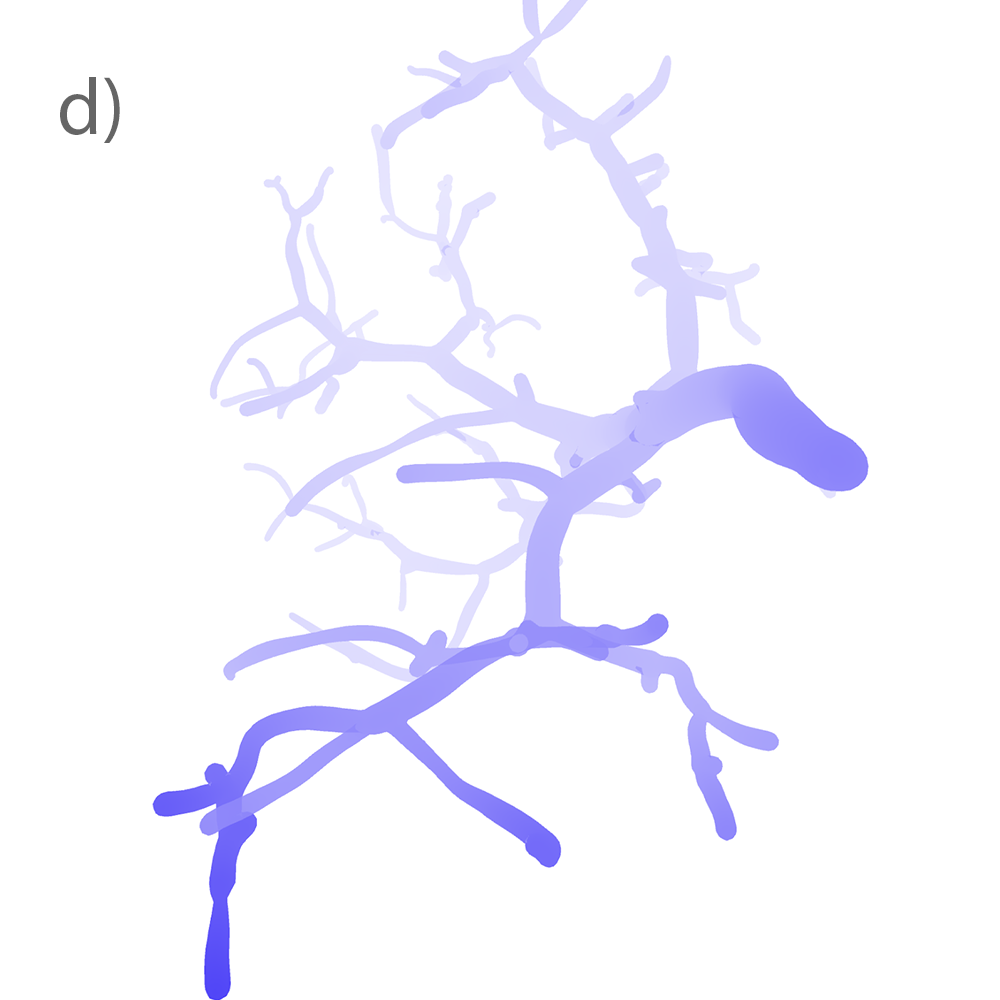}
\end{minipage}
\caption{The surface-based visualizations (a) heatmap, (b) isolines, and (c) pseudo-chromadepth, and (d) fog applied on a vascular structure.}
\label{fig:surface}
\end{figure*}
\section{Shading Fundamentals}
In this section, we briefly explain basic shading algorithms. 
Shading comprises diffuse, specular, and ambient reflection. There are highly realistic global and local illumination models as well as non-photorealistic shading techniques. When creating a non-photorealistic rendering, the specular part describes the reflection of light from smooth surfaces at certain angles, while the diffuse part is used to describe rough surfaces that reflect light in all directions. To create photorealistic renderings, many visualizations use the \textit{Fresnel term} to calculate realistic-looking reflectance.

\subsection{Phong}
\textit{Phong} shading
~\cite{Preim2016} is implemented as a base condition in numerous 
perception-based studies. 
The coloring of the surface is based on a combination of an opaque base color and 
shading effects resulting from a light source located somewhere in the scene, 
see Figure \ref{fig:fundamentals}\textcolor{blue}{a}.

\textbf{Implementation. }The \textit{Phong} visualization requires three input parameters: The mesh color, the shininess, and the specular color. Each parameter is passed to the shader as a shader property. The calculation of the ambient, specular, and diffuse components is performed within the fragment shader, taking into account the specified properties. 

\subsection{Toon}
\textit{Toon shading} (also called cel-shading) is a non-photorealistic surface 
rendering to mimic a cartoon representation~\cite{Decaudin1996}. 
Although there are several ways to implement toon shading, we will focus on one in particular in this paper.
Here, conventional smooth lighting values are calculated for each pixel and discretized 
into a small number of shades to create the characteristic "flat look". 
Thus, the shadows and highlights appear as blocks of color instead of being evenly
blended in a gradient, see Figure \ref{fig:fundamentals}\textcolor{blue}{b}.

\textbf{Implementation.} The \textit{Toon} visualization is based on a threshold calculation derived from the light calculation of the \textit{Phong} model. It requires seven input parameters: the mesh color, the shininess, the specular color, the ambient color, the rim color, the rim amount, and the rim threshold. Each of the parameters is passed to the shader as a shader property and can be customized to make the \textit{Toon} visualization unique. The important calculations of the ambient, specular, and diffuse components are performed in the fragment shader, similar to \textit{Phong} shading.

\subsection{Fresnel}
The \textit{Fresnel effect} describes the extent of reflection and refraction of light on 
a surface in relation to the viewing angle~\cite{Schlick1993}.
The flatter the viewing angle on a surface, the more light is reflected, resulting 
in the surface appearing brighter when illuminated, see Figure \ref{fig:fundamentals}\textcolor{blue}{c}.
A physically exact calculation of this effect is quite complicated, especially 
if one takes into account that the strength of the  \textit{Fresnel effect} also depends 
on the wavelengths of the light components due to chromatic dispersion.
Instead, we use a simplified version of this effect.

\textbf{Implementation. } The \textit{Fresnel} visualization requires three input parameters: The mesh color, the rim color (Fresnel color), and the Fresnel exponent. The Fresnel calculation is performed in a surface shader. By calculating the dot product between the surface normal and the viewing direction, the edge region can be determined. The edge region is then highlighted according to the Fresnel color and exponent.

%% file: chapter/07_surface_techniques.tex
\section{Surface-based Visualizations}
This section describes techniques that require surface data to convey additional information. These techniques modify the visual representation of the surface either partially in regions of high interest or completely to convey information along the entire mesh. Both egocentric and exocentric distances can be encoded this way.

\subsection{Heatmaps}
\textit{Heatmaps} are designed to convey information through color. They commonly indicate an area of varying interest.
While \textit{heatmaps} are typically used along with 2D data, an increasing number of studies are using \textit{heatmaps} as a visual tool to focus the user's attention on specific parts of 3D meshes.
In particular, \textit{heatmaps} are increasingly used in treatment planning and training in interventional radiology and surgery. 
In 2D \cite{kraus2020assessing}, \textit{heatmaps} can be used to represent quantitative data such as gaze information and city population.
In 3D, \textit{heatmaps} can be used to visualize, for example, the distances between vascularity and tumors \cite{hombeck2019evaluation,hombeck2022evaluating} or the degree to which the mitral valve is open\cite{eulzer2019temporal,lichtenberg2020mitral}. 
The color of the \textit{heatmap} then represents the quantitative data defined in the particular scenario, see Figure \ref{fig:surface}\textcolor{blue}{a}.
The color gradient must be defined by a colormap appropriate for the scenario. 
To ensure a smooth transition between the mesh and the \textit{heatmap}, one side of the color gradient is usually defined by the color of the underlying mesh.
If the mesh is colored according to the \textit{heatmap} one has to use shading carefully and assure that different lighting conditions do not impair the readability of the colormap.

\textbf{Implementation. } For this implementation, the \textit{heatmap} requires three parameters: the tumor position, the heatmap radius, and the color in the center of the \textit{heatmap}. While the heatmap radius and color are passed to the shader as shader properties, the heatmap position is updated as a uniform variable by a script. These parameters are used to calculate the distance between each vertex and the tumor position. If the vertex is within the heatmap radius, the color of that vertex is adjusted based on the interpolated color between the surface color and the heatmap color. The shader also allows for multiple tumor positions to be entered if multiple tumors are present. 

\subsection{Isolines} 
\textit{Isolines}, or contour lines, are lines that visualize data points from homogeneous regions within surface areas.
\textit{Isolines} help the user to understand the relationships between the surface and underlying acquired data, such as elevation or distance.
Common applications for \textit{isolines} are elevation maps for geographic data or atmospheric pressure maps.
By dividing the surface into several homogeneous regions, the number of annotations can be reduced to a minimum and a more structured illustration can be achieved.
This concept can also be applied to 3D mesh data. For example, \textit{isolines} can convey the distance information between the vascular structure and a tumor location, similar to \textit{heatmaps}~\cite{Tappenbeck:2006}, see Figure \ref{fig:surface}\textcolor{blue}{b}. 
Therefore, \textit{isolines} are represented as equidistant lines on top of the vessel.
In a 3D application, \textit{isolines} may continue outside the visible region, e.g., on the backside of the vascular structure, and reappear in different regions.
Matching these disappearing and reappearing \textit{isolines} can be challenging. 
To facilitate this process, the thickness of every other isoline should be adjusted.

\textbf{Implementation.} The \textit{isolines} require four input parameters: the tumor position, the isoline radius, the number of isolines, and the isoline thickness. While the isoline radius, number, and thickness are passed to the shader as a shader property, the tumor position is updated by a script. The isoline radius is divided into equidistant radii based on the number of isolines desired. The resulting radii are then thickened by the desired isoline thickness to create a tube at each radius. The thickness of every other isoline is calculated as thinner. Now the distance between each vertex and the tumor position is calculated. If the vertex is inside one of the tubes around the radii, the color of the vertex is set to black. The shader also allows you to enter multiple tumor positions. 

\subsection{Pseudo-Chromadepth}
Steenblick~\cite{Steenblik1987} introduced the \textit{chromadepth} technique, 
where depths is encoded by the visible color spectrum. 
However, this results in depth encodings based on a variety of hues, which can be 
distracting from the shading information. 
Since the shading is necessary to perceive the 3D structure, Ropinski et al.~\cite{Ropinski2006} 
introduced the \textit{pseudo-chromadepth} (PCD) technique, which uses a reduced 
number of hues to encode depth information. 
Inspired by the scattering of light in the atmosphere, a red-to-blue color scale 
is employed. 
Surface areas closest to the observer are shown in red, while a blue area 
signifies the greatest distance, see Figure \ref{fig:surface}\textcolor{blue}{c}.
Intervening colors are linearly interpolated based on depth values in the 
range $0$ to $1$, resulting in a color spectrum consisting of red, magenta, 
violet, and blue hues. 

\textbf{Implementation. } The \textit{PCD} requires one input parameter: The bounding box of the mesh. Although the PCD's color gradient is determined by the visualization itself, we provide the additional feature of specifying custom color maps by including the minimum and maximum distance color as a shader property. After calculating the minimum and maximum depth based on the bounding box of the mesh, the color of each fragment is determined by its depth within that volume. The color is interpolated between the maximum and minimum color values based on its depth value.

\subsection{Fog}
Gibson~\cite{Gibson1950} introduced a contrast-based depth encoding, 
called \textit{air perspective}. Depth perception depends on the scattering 
of light in the atmosphere.
The basic concept is that distant objects scatter incident light more before the light reaches the eye than nearby objects. 
Thus, distant objects are perceived with less contrast.  
Kersten-Oertel et al.~\cite{Kersten-Oertel14} modelled the aerial perspective 
as a fog-like effect by reducing the contrast with increasing distance from the camera. 
The color spectrum consists of a single color, where the corresponding alpha 
value is changed from opaque to transparent based on the depth value.
With this, nearer and farther parts appear opaque and transparent, respectively, 
as shown in Figure \ref{fig:surface}\textcolor{blue}{d}.
Thus, this technique is similar to \textit{PCD} shading in terms of encoding 
depth information based on egocentric distances. 
However, instead of mapping depth information to colors, it is encoded using 
transparency.

\textbf{Implementation.} The \textit{fog visualization} requires one input parameter: The transparency falloff. For each fragment, we calculate its depth value and determine the transparency based on this depth and the transparency falloff parameter. The higher the falloff value, the faster the object fades into the fog.  

\subsection{Stereoscopic Vision}
Stereopsis is an important factor influencing depth perception. It describes the process of viewing the environment with both eyes. Instead of processing only one image, two slightly shifted images are processed by our brain. These binocular disparities are then used to create depth information. While in computer graphics we try to fake depth by using perspective renderings, the results are still displayed as a single image. Rendering stereoscopic images with 3D glasses or HMDs improves depth perception in almost all cases compared to their desktop counterpart~\cite{hombeck2022evaluating}. Thus, using stereoscopic renderings and devices can help to better perceive the scenario and further understand the depth information. Stereoscopic rendering can be combined with all presented visualizations.

%% file: chapter/08_auxiliary_tools.tex
\begin{figure*}
\centering
\begin{minipage}{.33\textwidth}
  \centering
  \includegraphics[width=\linewidth]{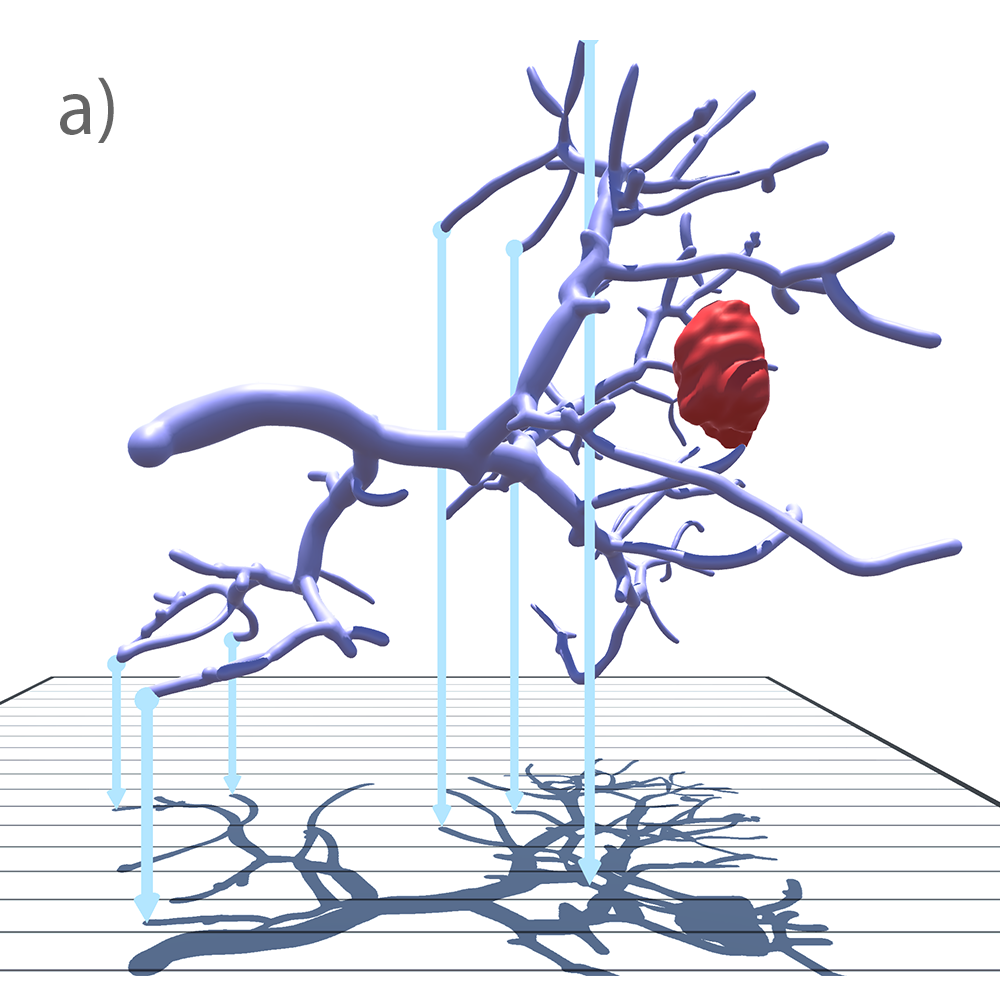}
  \label{fig:test1}
\end{minipage}%
\begin{minipage}{.33\textwidth}
  \centering
  \includegraphics[width=\linewidth]{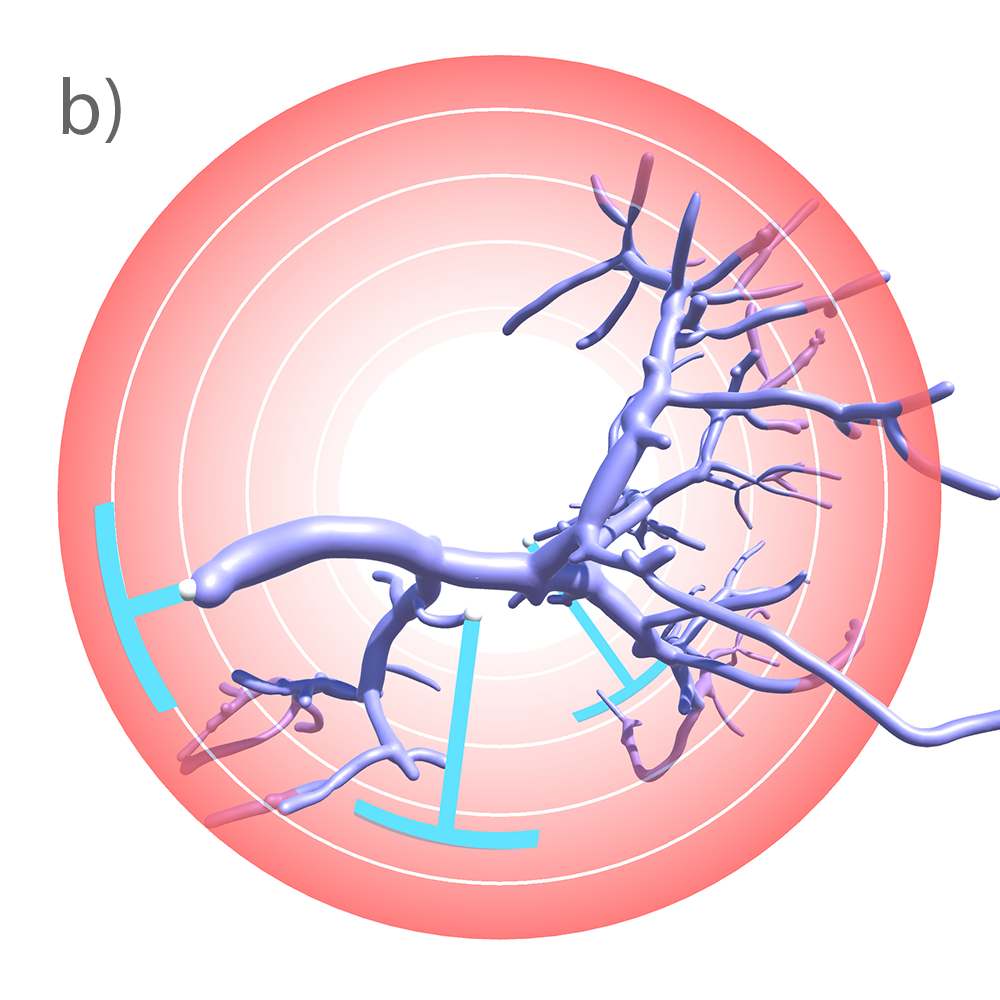}
  \label{fig:test2}
\end{minipage}
\begin{minipage}{.33\textwidth}
  \centering
  \includegraphics[width=\linewidth]{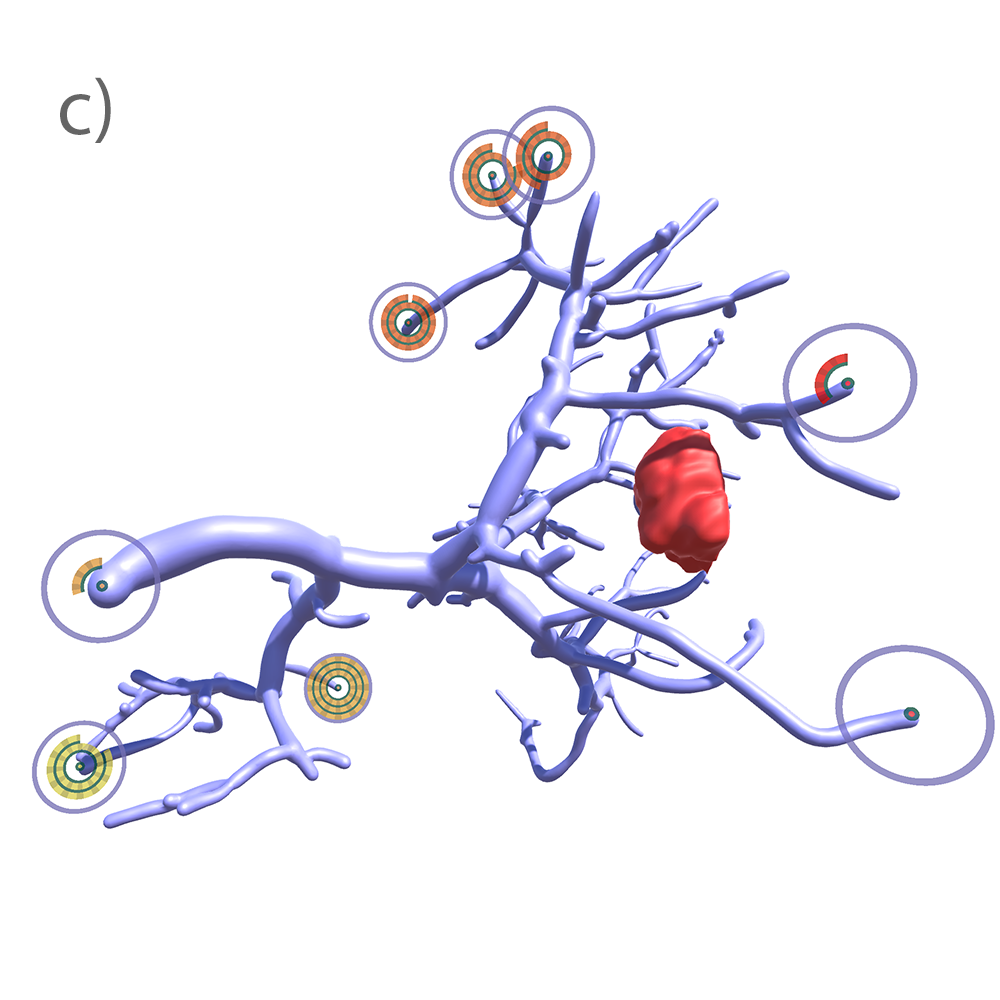}
  \label{fig:test2}
  
\end{minipage}
\begin{minipage}{.33\textwidth}
  \centering
  \includegraphics[width=\linewidth]{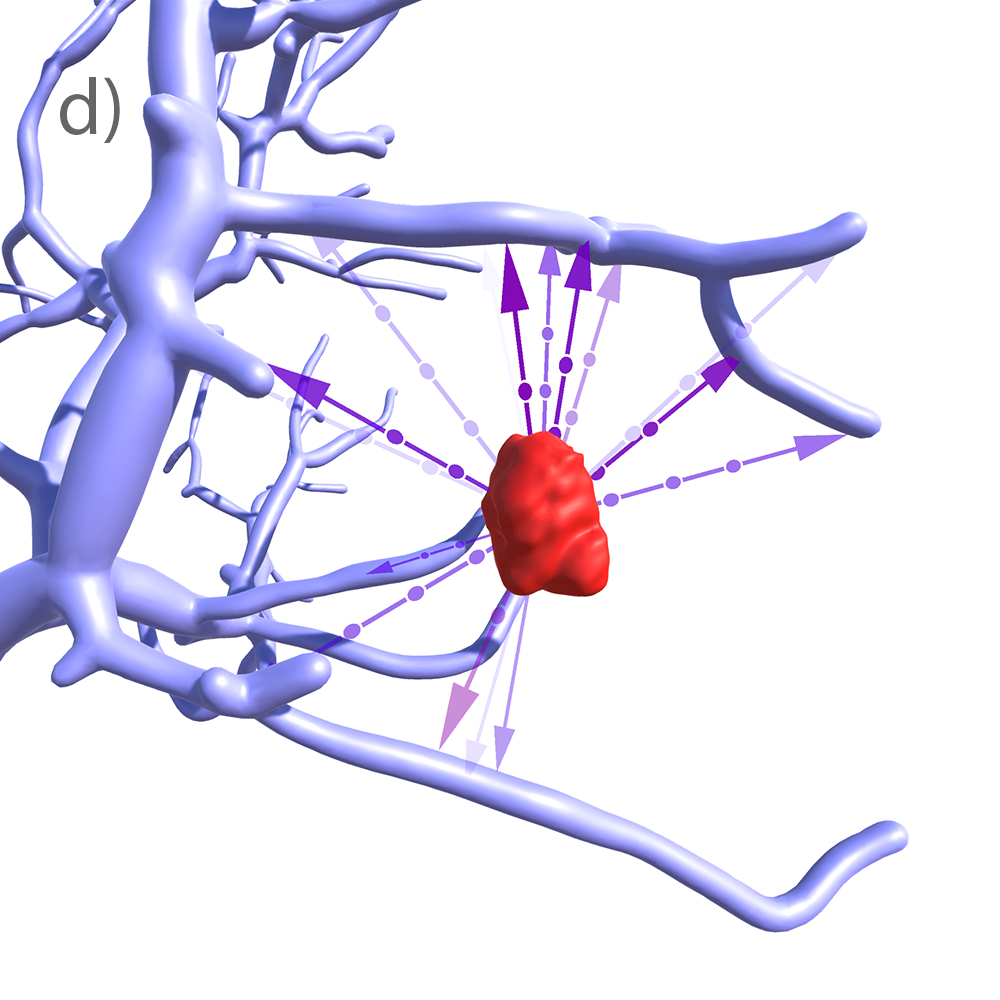}
  \label{fig:test2}
\end{minipage}
\begin{minipage}{.33\textwidth}
  \centering
  \includegraphics[width=\linewidth]{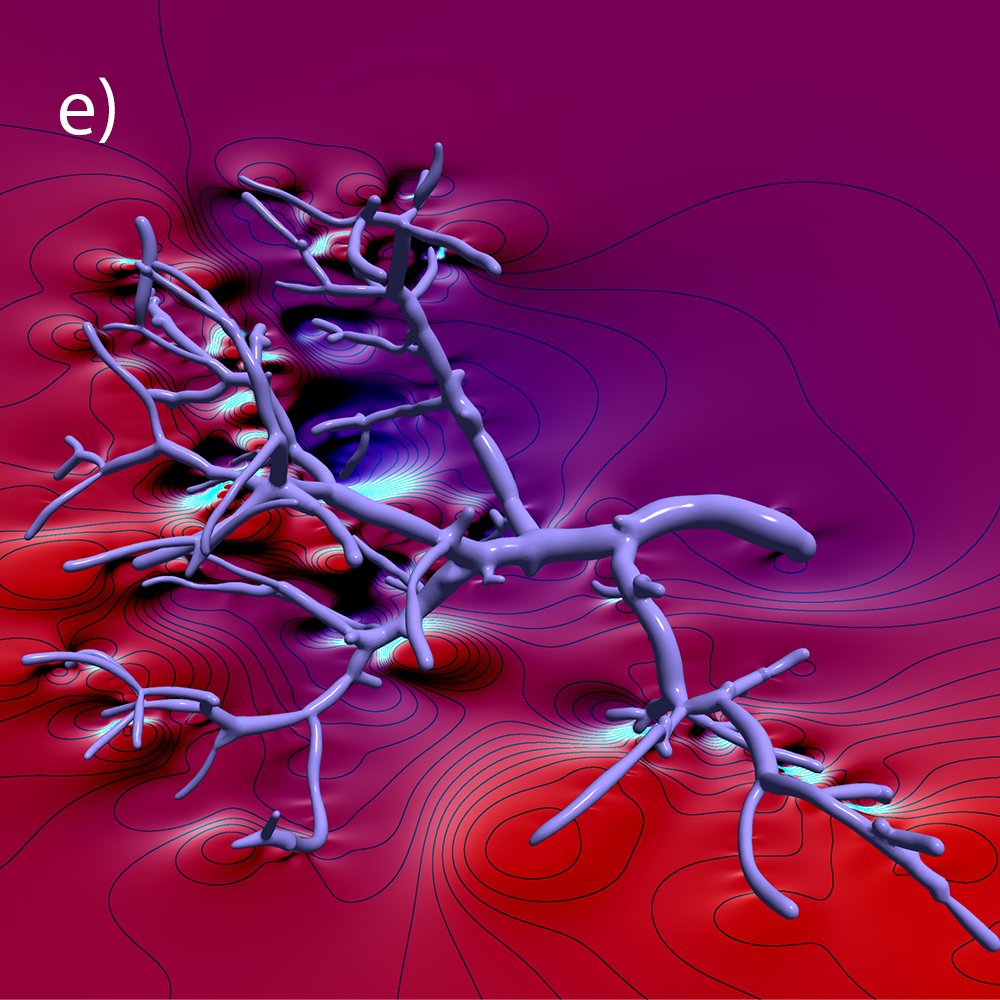}
  \label{fig:test2}
\end{minipage}
\begin{minipage}{.33\textwidth}
  \centering
  \includegraphics[width=\linewidth]{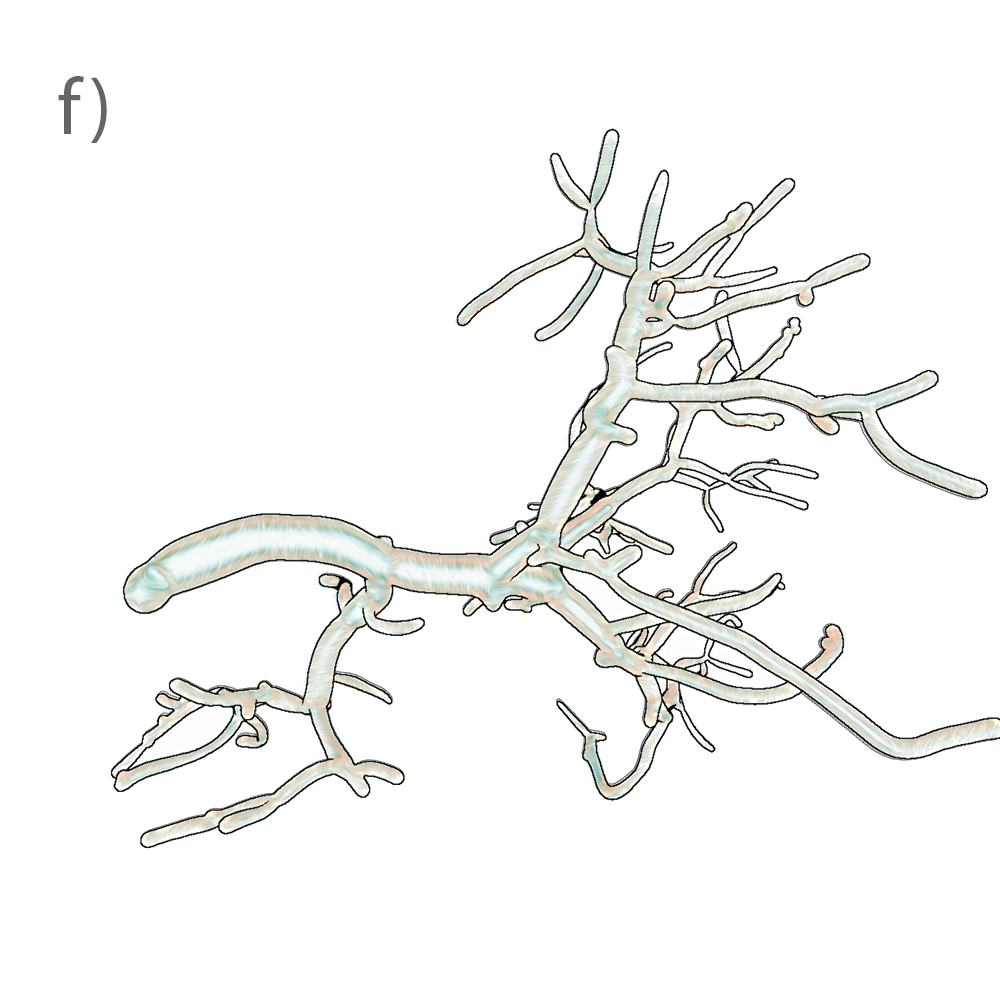}
  \label{fig:test2}
\end{minipage}


\begin{minipage}{.33\textwidth}
  \centering
  \includegraphics[width=\linewidth]{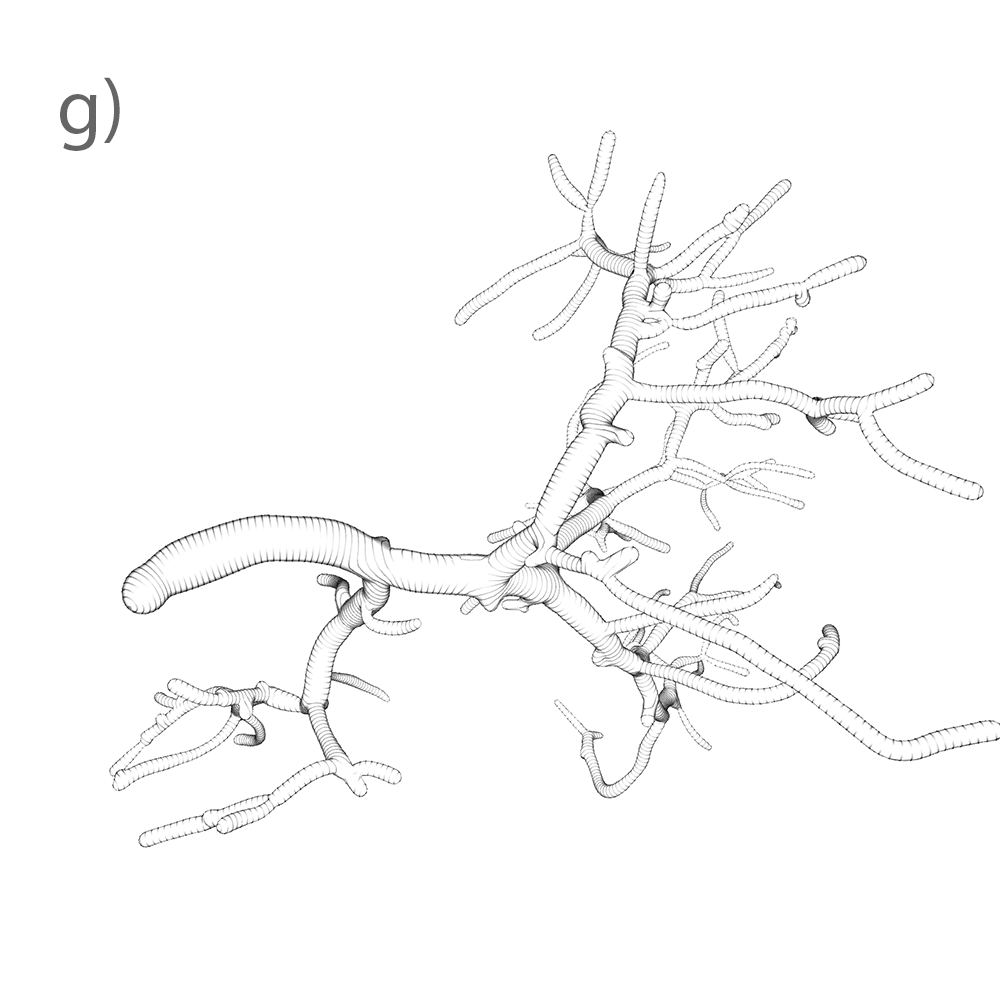}
  \label{fig:test2}
\end{minipage}
\begin{minipage}{.33\textwidth}
  \centering
  \includegraphics[width=\linewidth]{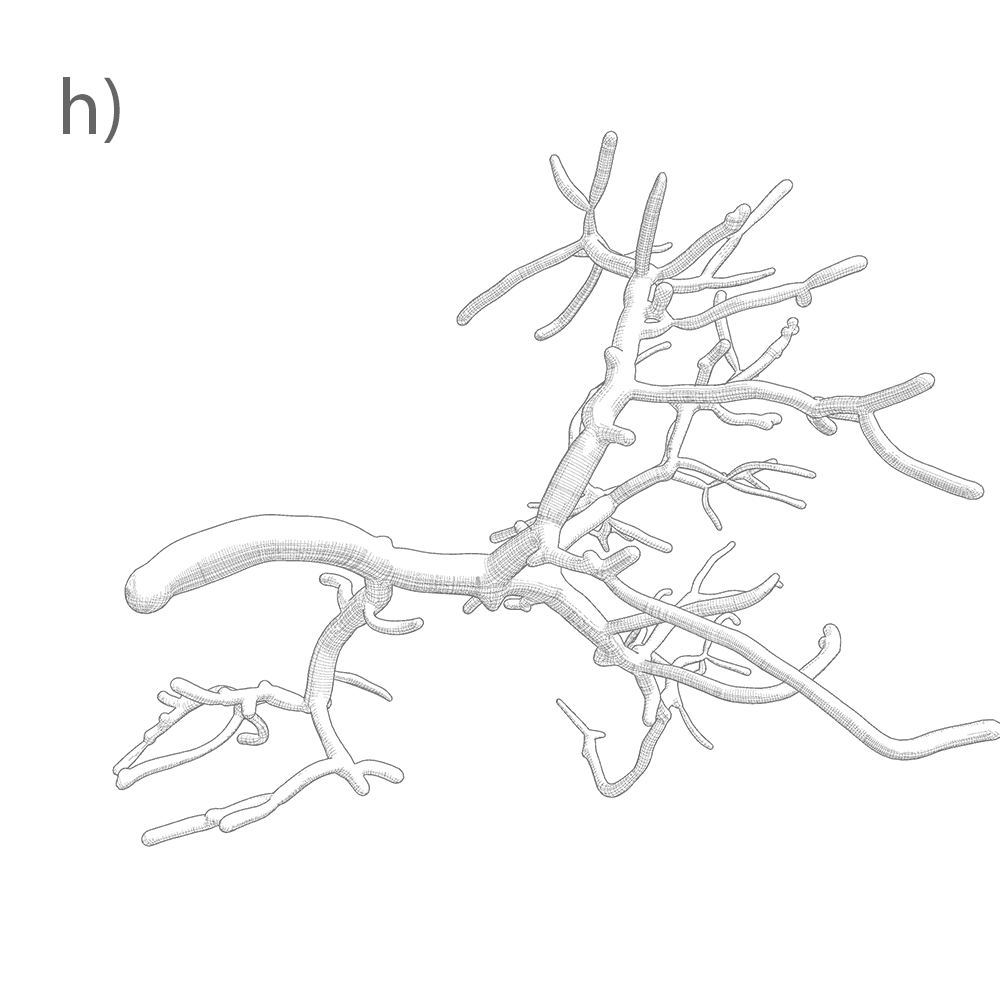}
  \label{fig:test2}
\end{minipage}
\caption{Visualizing a Vascular Structure: Using (a) Supporting Lines, (b) Supporting Anchors, (c) Concentric Circle Glyphs, (d) Arrow Glyphs, (e) Void Space Surfaces, (f)  Mesh Curvature as Vector Field with LIC, (g) Hatching, and (h) Hatching by Hertzmann and Zorin.}
\label{fig:auxiliary}
\end{figure*}

\section{Auxiliary Tools}
Here we describe a category of perception-supporting visualization techniques that have been identified by Preim et al.~\cite{Preim2016} as \textit{Auxiliary Tools}. 
The following is mainly a recap of the publication by Lichtenberg and Lawonn~\cite{rea_auxiliary_2019}, providing an overview of existing \textit{Auxiliary Tools} for enhanced depth perception in the context of vascular visualization and defining the term as:
\textit{"Auxiliary Tools in-depth perception describe visual entities, (i.e., geometric
objects) that augment a generated image of spatial data in order to encode
depth information or to trigger and/or exaggerate depth cues."}
Visual information channels on the surface of interest are therefore not directly affected, while additional depth information is packed into the final visualization.
However, the additional geometry may overlap with the medical data, leading to a cluttered view and impaired perception of some parts of the visualization.
While glyphs explicitly encode some given information in an arbitrary way they may become an \textit{Auxiliary Tool} for depth perception.
This is the case if they are designed and work as a trigger for depth cues and therefore encode depth information.
The design of an \textit{Auxiliary Tool} should always consider the pre-attentive and attentive phase of human perception.
\subsection{Supporting Lines}
The \textit{Supporting Lines} (SL) by Lawonn et al.~\cite{Lawonn_2015_MICCAI} utilize the shadow of an object as a natural depth cue. 
A vessel model is located above a plane with a shadow projection of the model, see Figure~\ref{fig:auxiliary}\textcolor{blue}{a}. 
On the shadow plane, it is easy to estimate which point is further away from the viewer.
The projection can by stylized to highlight certain parts of the model, such as tumor or vessel tissue.
\textit{SL} can be added by the user by selecting points on the vasculature. 
These points are then connected to their corresponding location on the shadow projection.
Thus, a surface point and its shadow have an unambiguous connection which is useful in otherwise unclear cases.
\textit{SL} further invoke an overlapping depth cue with other parts of the vessel geometry. 
It enables the user to gain additional depth information for whole vessel branches with respect to a selected point.

\textbf{Implementation. } The shadow projection is achieved by a directional light source that is perpendicular to the ground plane. In this way, an orthographic shadow is projected onto the shadow plane. The supporting lines are generated by specifying a starting point on the vessel surface. From there, the supporting line geometry is created and scaled to the shadow plane by a script.

\subsection{Supporting Anchors}
A follow-up technique to the one above are the \textit{Supporting Anchors} (SA) by Lawonn et al.~\cite{Lawonn:2017:ISP:3067926.3067940}.
A drawback of the previous method is that the shadow plane is located at a fixed location outside the vascular model.
The \textit{SA}, however, are connected to an open cylinder which is view-centered and oriented along the view direction, see Figure~\ref{fig:auxiliary}\textcolor{blue}{b}. 
In this way, the selected points can be connected to their reference surface in any direction by projecting them to the closest point on the cylinder.
This reduces visual clutter and allows to use the cylinder as an intuitive probe to select and highlight an area of interest within the vasculature.
The connections then are augmented with an anchor-like extension along the cylinder within the same depth plane. 
This simplifies the comparison of reference points that are further away from each other.

The selection of points is furthermore done automatically, using an algorithm that picks end points of the vessel structure with respect to their spatial distribution inside the cylinder.
Overlapping anchor geometry is therefore reduced.

\textbf{Implementation. }
For the \textit{Supporting Anchors}, a cylindrical object is generated and aligned with the camera position. This cylinder is rendered using a combination of fog and isoline visualization to create the desired transparency falloff with supporting lines. The \textit{SA} are created in a similar manner to the supporting lines. After selecting a position on the vascular structure, the anchor geometry is created and scaled by a script. 

\subsection{Concentric Circle Glyphs}
The \textit{Concentric Circle Glyphs} (CCG) by Lichtenberg et al.~\cite{LichtenbergHL17}
are disc-shaped glyphs and also located at vessel endpoints and automatically selected by an algorithm to reduce overlap.
Distance to the viewer is 
encoded by three concentric circles that get filled in sequence from inside out and in a clockwise manner, see Figure~\ref{fig:auxiliary}\textcolor{blue}{c}.
Three full circles are equal to the maximal distance and no circles at all are equal to the minimal distance.
Additionally, the glyph size depends on the distance to the viewer.
Glyphs further away are smaller in world space. 
The color of the glyphs can be used to indicate the proximity of the glyph relative to the tumor. 
In this way, the \textit{CCG} can be interpreted during the pre-attentive phase and act as an \textit{Auxiliary Tool}. 

\textbf{Implementation. }
For each \textit{Concentric Circle Glyph}, the distance to the camera and the distance to the tumor is required and passed to the shader via a script. The fullness and color of each circle are calculated in the fragment shader based on these distance values. For each glyph, a view-aligned quad is created and rendered using the concentric circle glyph shader.

\subsection{Arrow Glyphs}
\textit{Arrow glyphs} are a combination of color \cite{Levkowitz1991} and shape \cite{Wittenbrink1996} that can be used to encode distances. Unlike surface visualizations, arrow glyphs do not utilize the surface as canvas, but instead, use the 3D space between the reference point and the surface. The visualization consists of multiple arrows pointing towards the surface, where the distribution of the arrows depends on the chosen endpoint, see Figure~\ref{fig:auxiliary}\textcolor{blue}{d}. While an equidistant distribution is not naturally given, it can be computed by an algorithm of Lichtenberg et al. \cite{NL18}. Given these sample points, any desired density of glyphs can be achieved.

Hombeck et al. \cite{hombeck2019evaluation} propose two density levels for arrow glyphs: a denser distribution in critical areas, e.g., near vasculature, and a less dense distribution in non-critical areas. Adjusting the density can increase the information gain in critical areas while maintaining a more structured view in non-critical areas. In addition to arrow shape, size, color, and transparency are used to encode information about the distance and shape of the mesh. The color of the arrow changes within hues of red, magenta, purple, and blue, similar to the gradient of the pseudo-chromadepth visualization. To provide information about the relative distances, each arrow contains a small dot every 2 cm. As the number of glyphs is increased, the size of the arrow is reduced to avoid cluttering the view. In addition to distance encoding, the arrow glyphs can also encode shape information through transparency. The larger the angle between the arrow and the surface normal, the more transparently the glyph is rendered. 

\textbf{Implementation.} \textit{Arrow glyphs} require five input parameters: the tumor position, the maximum length, the switching distance, the thickness, and the glyph texture. While the maximum length, the switching distance, and the glyph texture are passed to the shader as shader properties, the tumor position and whether the mesh is considered a large or small glyph are updated by a script. While the mesh vertices can serve as anchor points for the glyphs, additional sampling of points across the surface is required for an equidistant distribution and is already provided in the example. For each input vertex, the geometry shader is applied and generates a view-aligned quad between the input vertex and the tumor position. This square is filled with the selected glyph texture. The color and transparency are then calculated in the fragment shader. The shader also allows the input of multiple tumor positions. 

\subsection{Void Space Surfaces}
The \textit{Void Space Surfaces} by Kreiser et al.~\cite{kreiser2018void} are a plain screen space method to convey depth information, 
%
where no actual additional geometry is used.
Instead, the space in a rendered image that is not occupied by the projection of the displayed vessel structure is filled with a smooth surface that attaches to the contours of the vasculature, see Figure~\ref{fig:auxiliary}\textcolor{blue}{e}.
Shading, pseudo-chromadepth, and iso-lines support the pre-attentive and attentive depth perception.
The method does not require any selection of surface points to highlight and is conceptually free of any overlaps. 
On the downside, the depths of point pairs that are not in close proximity in screen space are not easy to compare.

\textbf{Implementation. }
The normalized and linearized depth texture of the camera is required for the representation of \textit{Void Space Surfaces}.
After obtaining the depth texture, the contour lines of the vessel object are extracted and stored in a texture.
Each point on the contour line is then passed to the shader and used as a weight to fill the void space. 
At the end of this calculation, the illuminance, isolines, and colors are calculated and stored in another texture.
The output texture is set as the background behind the vessel object.


%% file: chapter/09_illustrative.tex
\section{Illustrative Visualization}
Illustrative visualization is inspired by traditional illustration, e.g., in textbooks, and can be used to support the perception of depth and shape~\cite{Lawonn:2018:SSI}.
To convey shape or depth, \emph{hatching methods} are commonly used. 
Hatching is an illustration technique, where the object is covered with a dense set of lines.
Originally, the density of lines was used to shade the object, i.e., regions that are brighter are depicted with fewer lines than dark-shaded regions, where a lot of lines are used.

Ritter et al.~\cite{Ritter2006} varied the hatching strokes to encode the distance to the viewer.
Vessel segments that are close to the viewer are illustrated with thicker black lines and regions that are far away are illustrated with thinner lines.
For illustration purposes, Lawonn et al.~\cite{Lawonn_2015_MICCAI, Lawonn:2017:ISP:3067926.3067940} employed hatching lines based on the \emph{Confis} method~\cite{Lawonn_2013_CGF} and their extension~\cite{Lawonn_2014_CGF} to convey the shape of the vessel structure, see Figure~\ref{fig:auxiliary}\textcolor{blue}{g}.
To encode the depth, they introduced glyph-based visualizations: \emph{supporting lines} and \emph{supporting anchors}.
One limitation of the first approach~\cite{Lawonn_2015_MICCAI} was that the position of the glyph placement was fixed and needed to be set in advance.
This limitation was then resolved by automatically placing the glyphs based on the view position~\cite{Lawonn:2017:ISP:3067926.3067940}.
Later, Lichtenberg et al.\cite{LichtenbergHL17} extended this idea and placed the glyphs based on a cost function.
Again, they employed hatching strokes to convey the shape of the vessel structures.
Instead of pure illustration-based hatching strokes, Lichtenberg et al.~\cite{NL18} employed a real-time stripe pattern approach.
This technique allows to vary the stroke size based on an arbitrary scalar field defined on the surface, e.g., curvature. 
It was also possible to alter the thickness of the strokes in real-time to encode the distance to the viewer or any arbitrary object.
Hatching techniques allow an encoding on the surface, while the surface can still be colored to encode additional information~\cite{Meuschke_2017_TVCG, Smit2017}.
While we implement the supporting lines, supporting anchors, and hatching, more illustrative visualizations will follow as further work.

\subsection{Hatching}
To optimize performance, we offer two distinct approaches for hatching. The first approach is real-time but requires preprocessing outside of the framework. The second approach is based on the hatching by Hertzmann and Zorin \cite{hertzmann2000illustrating} and can be applied to any surface, but is semi-real-time. Although the visualization can be displayed in real-time, it must be recalculated when the perspective changes. 

For the first approach, a hatching-based mesh is created outside the Unity framework due to its limited mesh creation, editing, and saving capabilities. The mesh is expanded by 1mm along the normal direction to prevent buffer issues. The generated mesh displays the hatching pattern for a single object and is overlaid on the original mesh. The hatching is rendered so that it is visible only at the edges of the vascular structure and fades to transparency in the center.  Precomputing and incorporating the hatching lines into Unity has both advantages and disadvantages. On the one hand, it enables real-time rendering of a computationally intensive visualization, but on the other hand, it lacks the flexibility to make adjustments, such as changing the hatching line thickness after mesh generation.

\textbf{Implementation}
The Unity scene comprises two separate meshes: the original mesh and the hatching mesh, with the latter overlaid on top of the former. The hatching transparency is controlled by a transparency shader applied specifically to the hatching mesh, providing the ability to independently modify the visualization of the original mesh as desired.

\subsection{Hatching by Hertzmann and Zorin}
To increase the hatching flexibility, we incorporate an adapted version of the Hertzmann and Zorin \cite{hertzmann2000illustrating} hatching approach. see Figure \ref{fig:auxiliary}\textcolor{blue}{h}.
They draw inspiration from traditional line drawing techniques in order to effectively convey depth and structure. The proposed algorithm includes a method for rapidly calculating mesh contours using a dual homogeneous space and octrees. Additionally, the algorithm addresses the generation of hatching lines through the use of cross fields, defined as an unordered pair of perpendicular directions on a surface, initialized with principal curvatures. An optimization problem is solved to align cross fields in poorly defined regions with neighboring well-defined regions. The hatching lines are generated in image space through the use of a streamline algorithm, and selective reduction is employed to create tonal variations. 

For our work, the mesh contour was calculated using a GPU-based geometry shader rather than CPU-based algorithms. This improvement of the algorithm was only possible of the advancements in graphics hardware since the publication of the paper. The optimization of the crossfield was performed using the limited-memory Broyden–Fletcher–Goldfarb–Shanno (L-BFGS) algorithm and the \textbf{ALGLIB} library \cite{ALGLIBCC59:online}, a cross-platform numerical analysis and data processing library. This results in a collection of angles with respect to the principal curvature direction, which are then interpolated on the mesh and projected into screen space. Our work employs a C\# Sharp implementation of the curvature estimation, as presented by Rusinkiewicz \cite{rusinkiewicz2004estimating}, which replaces the original algorithm for calculating mesh curvatures through subdivisions. An evenly spaced streamlines algorithm \cite{jobard1997creating} was utilized to sample the crossfield image and generate the hatching lines. To ensure alignment during the reduction of hatches, a breadth-first search (BFS) was employed. A geometry shader was written to render the resulting hatches and contours as a single mesh.

\textbf{Implementation}
Within our framework, the hatching script begins by adding and initializing all the necessary scripts. During initialization, curvatures are calculated and the crossfield is optimized once. The first step in calculating the hatching involves finding the contours of an object, determining self-intersections, and eliminating occluded contours. Crossfield directions are then projected and interpolated using a geometry shader. Then a separate image is rendered for brightness values used for hatching reduction. The resulting hatching lines are created as streamlines and pruned as needed. The output is a mesh with contours, hatch lines, and width information based on brightness, rendered by a custom line shader. Hatches are updated only when parameters change or the camera or mesh are moved. 
The streamlines algorithm for generating evenly spaced hatching has a high computational demand as it calculates in screen space. This also leads to a lack of frame coherence, as the streamlines are sensitive to even small variations in starting conditions. These issues can be improved by performing the streamline calculation in texture space, as suggested by Spencer \cite{spencer2009evenly}.

%% file: chapter/10a_LIC.tex
\section{Surface Fields}
In the following section, we will explore scalar and vector fields in detail. We will discuss their use in representing physical properties in biological systems, their advantages, and how they can be visualized to gain a deeper understanding of the underlying physics. By understanding the role of scalar and vector fields in the modeling of biological systems, we can gain valuable insight into the complex interactions and relationships between various physical properties.

\begin{figure}
    \centering
    \includegraphics[width=0.7\linewidth]{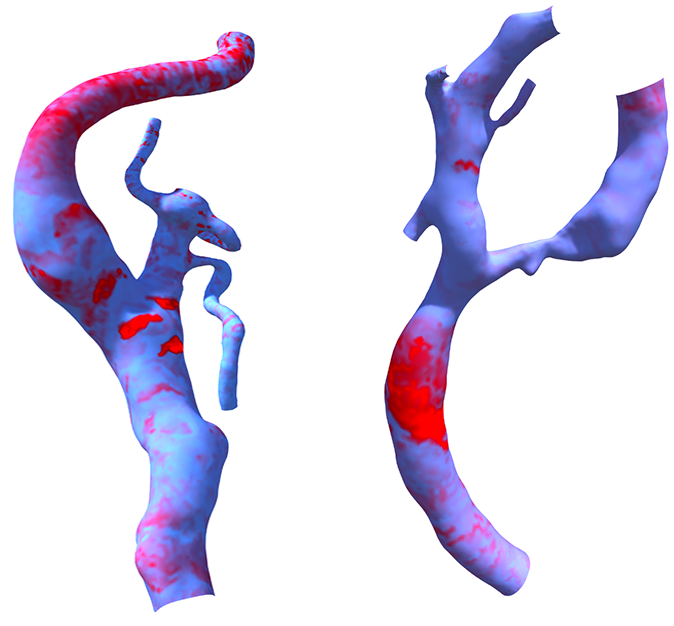}
    \caption{Two examples of wall shear stress, represented as a scalar field superimposed on a mesh surface, demonstrating the force exerted by blood on the inner wall.}
    \label{fig:scalarfield}
\end{figure}
\subsection{Scalar Field}

In computational modeling of biological systems, scalar fields \cite{yan2021scalar} play a crucial role in representing physical quantities, such as pressure, flow, and concentration of different substances. These scalar fields are utilized in various applications, including the modeling of blood flow in the circulatory system. The accurate representation of the flow of blood through these networks is essential in order to comprehend and predict physiological processes, such as the delivery of oxygen to tissues and the removal of waste products.

In order to model the circulatory system accurately, it is imperative to represent the physical properties of the fluid, such as pressure and velocity, in a manner that accurately reflects the underlying physics. Scalar fields serve as an efficient and versatile means of representing these properties visually. 
The ability to visualize scalar fields is valuable for researchers and medical professionals for a deeper understanding of the underlying blood flow. 

\textbf{Implementation}
We generate vector fields by assigning a scalar value to each vertex. This scalar value can be either a sorted list of values or a mapping of vertex ID to scalar value. The scalars are then mapped to a color map, which can be selected from a predefined set or created as a custom map. The final color is calculated using a fragment shader that performs the color mapping.

\subsection{Vector Fields }

An advanced form of visualizing physical properties in biological systems is the representation of vector fields. This type of visualization provides a comprehensive understanding of the direction and magnitude of physical quantities such as flow, velocity, and force. The combination of scalar and vector field visualization provides a powerful tool for comprehending the complex interactions and relationships between different physical properties in biological systems.

The Line Integral Convolution (LIC) is a computational technique that effectively visualizes such vector fields by integrating the field along streamlines, which are lines that represent the direction of flow at each point in the system, see Figure \ref{fig:auxiliary}\textcolor{blue}{f}.

LIC begins by initializing a noise texture, typically a random image, and then integrating the vector field along the calculated streamlines. The noise texture is advected along the streamlines, creating a coherent image that represents the vector field \cite{cabral1993imaging}. The intensity of the noise at each pixel can be used to denote the magnitude of the field.
LIC is a valuable tool for visualizing the underlying structure of complex and dynamic systems, particularly in medical imaging and scientific visualization. In these domains, it is used to analyze curvature and blood flow in vascular networks, where it is capable of revealing details such as flow direction and strength that are not visible with other visualization methods. Additionally, LIC can be seamlessly combined with other visualization techniques, such as volume rendering or surface-based visualization, to provide a more comprehensive representation of the system under investigation. The versatility of LIC makes it an ideal solution for visualizing complex data, either as a complementary tool to existing visualizations or as a standalone visualization approach.

\textbf{Implementation}
We utilize the pipeline developed by Lawonn et al. \cite{lawonn2014line}.
The pipeline contains several post processing steps: first, the screen space ambient occlusion (SSAO) of a given mesh is calculated and stored in a texture. 
The illumination gradient (IG) is used as the feature vector field, obtained by deriving the diffuse illumination of the object and stored in another texture. 
Then, each image pixel is transformed by a 2D look up table (LUT), where the SSAO value correspondents to the x-axis and the IG length to the y-axis.
The LUT can be used to weight the visual appearance of the SSAO and IG, e.g. through color and brightness.
The resulting image is multiplied with a noise image.
This noise image is generated by assigning a noise texture to an object and multiplying the brightness with the diffuse illumination to gain more visual depth.
In the last step, a LIC-shader is applied to the image.
An edge detection step prevents the screen space streamlines to bleed over edges.
The pipeline is implemented as a C\# script and attached to the camera.
In the script, various adjustments can be made for each pipeline step.
Objects that need to be drawn must have attached the given normal and noise material.

%% file: chapter/10b_Enpoints.tex
\begin{figure}
    \centering
    \includegraphics[width=0.7\linewidth]{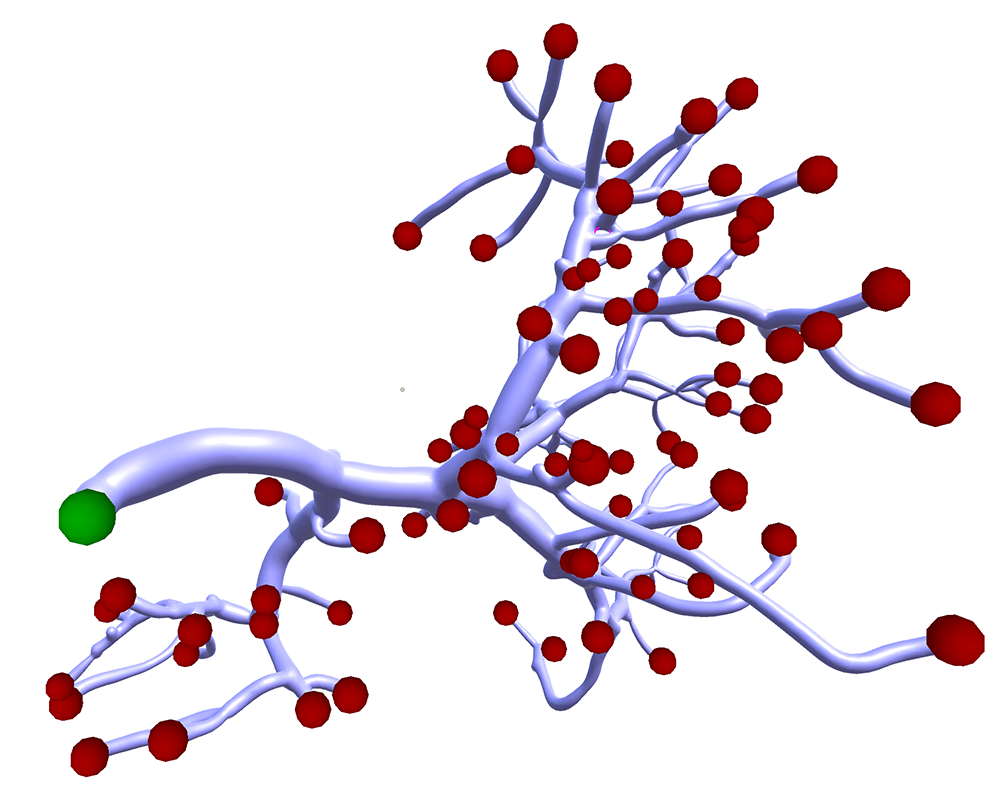}
    \caption{ Visualizing endpoint localization on a vascular model. Spheres represent the ends of each branch. }
    \label{fig:endpoints}
\end{figure}

\newcommand{\cmark}{{\ding{51}}}%
\newcommand{\xmark}{\textcolor{lightgray}{\ding{55}}}%
\newcommand{\omark}{$\odot$}%

	
\definecolor{Gray}{gray}{0.95}
\begin{table*} [t]

\caption{Overview of fundamental, surface-based, auxiliary, and illustrative visualizations. For each visualization, we indicate the occupied visual cues, the attentive phase, the distance encoding, and system information. Since perspective projection and partial occlusion are included in all visualizations, we have not explicitly mentioned them in the table.}
\label{tab:overview}
\adjustbox{max width=\textwidth}{
\begin{tabular}{||l|| c c c c c c c c c c c c c|} 
 \hline
 Visualization & Shading & Shadow & Color & Transparency &Surface & Void Space& Pre-attentive& Attentive & Egocentric & Exocentric & Realtime 
\\ [0.5ex] 
\rowcolor{Gray} 
\hline\hline Phong 
    & \cmark & \cmark & \xmark & \xmark & \xmark & \xmark & \cmark & \xmark & \cmark & \xmark & \cmark\\ 

\hline Toon 
    & \cmark & \cmark & \xmark & \xmark & \xmark & \xmark & \cmark & \xmark & \cmark & \xmark & \cmark\\
   \rowcolor{Gray}  
\hline Fresnel 
    & \cmark & \cmark & \xmark & \xmark & \xmark & \xmark & \cmark & \xmark & \cmark & \xmark & \cmark\\
    
\hline Supporting Lines 
    & \cmark & \cmark & \xmark & \xmark & \xmark & \cmark & \xmark & \cmark & \cmark & \cmark & \cmark\\
  \rowcolor{Gray}   
\hline Supporting Anchors 
    & \cmark & \xmark & \xmark & \omark & \xmark & \cmark & \xmark & \cmark & \cmark & \cmark & \cmark\\ 
    
\hline Concentric Circle Glpyhs 
    & \cmark & \xmark & \cmark & \xmark & \xmark & \omark & \xmark & \cmark & \cmark & \cmark & \cmark\\
    \rowcolor{Gray} 
\hline Void Space Surfaces 
    & \cmark & \xmark & \cmark & \xmark & \xmark & \cmark & \xmark & \cmark & \cmark & \xmark & \cmark\\ 
    
\hline Arrow Glyphs
    & \xmark & \xmark & \cmark & \cmark & \xmark & \cmark & \cmark & \cmark & \xmark & \cmark & \cmark\\ 
    \rowcolor{Gray} 
\hline Heatmaps 
    & \xmark & \xmark & \cmark & \xmark & \cmark & \xmark & \cmark & \cmark & \xmark & \cmark & \cmark\\ 
    
\hline Isolines 
    & \xmark & \xmark & \xmark & \xmark & \cmark & \xmark & \cmark & \cmark & \xmark & \cmark & \cmark\\
   \rowcolor{Gray}  
\hline Pseudo-Chromadepth 
    & \cmark & \xmark & \cmark & \xmark & \cmark & \xmark & \cmark & \xmark & \cmark & \xmark & \cmark\\
    
\hline Fog 
    & \cmark & \xmark & \xmark & \cmark & \cmark & \xmark & \cmark & \xmark & \cmark & \xmark & \cmark\\
  \rowcolor{Gray}   
\hline Hatching 
    & \xmark & \xmark & \xmark & \xmark & \xmark & \xmark & \cmark & \xmark & \cmark & \xmark & \cmark\\
\hline Hatching by H. and Z. 
    & \xmark & \cmark & \xmark & \xmark & \xmark & \xmark & \cmark & \xmark & \cmark & \xmark & \omark\\
\hline Scalar field 
    & \xmark & \xmark & \cmark & \xmark & \cmark & \xmark & \cmark & \cmark & \omark & \omark & \cmark\\
\hline Vector fields 
    & \xmark & \cmark & \xmark & \xmark & \cmark & \xmark & \cmark & \cmark & \omark & \omark & \cmark\\

\hline

\end{tabular}

}

\end{table*}
\section{Endpoint localization}
We extend our previous work by incorporating an algorithm to identify the specific points on the surface required for glyph placement such as anchor glyphs or concentric circle glyphs. In the case of vascular structures, these points are typically the endpoints of the structure.
Localization of endpoints plays a crucial role as the root point in vessel trees holds a unique significance. This endpoint identification can subsequently also be incorporated into more advanced visualization techniques. Here an algorithm by Lichtenberg and Lawonn~\cite{Lichtenberg2018_VCBM,lichtenberg2020parameterization} was implemented with optimizations by Cech~\cite{cech_2020}.

The endpoint localization is carried out in two stages. Firstly, the mesh undergoes iterative smoothing by relocating vertices using an implicit Laplacian, while preserving the edges. The updated vertex positions, represented as $p'$, can be solved by the equation:
\begin{equation}
{W_L L\choose W_H} p' = {0 \choose W_H p}
\label{eq:thinning}
\end{equation}
Where $W_L$ and $W_H$ are weight matrices that regulate smoothing and reduce smoothing intensity with each iteration. The smoothing process is terminated when the volume reaches near-zero or when the relative error exceeds 1\%.
Once the skeletonization is completed, the endpoint criterion is evaluated for each vertex. A vertex is considered an endpoint if all other vertices within its vicinity lie in the same hemisphere. Subsequently, the root point is identified by determining the endpoint in the original mesh with the lowest absolute mean curvature.

The calculation of the mathematical problem can be time-intensive, especially for large meshes. Our tests show that it took approximately 2 minutes to perform the calculation on a mesh with roughly 30,000 vertices. Fortunately, the calculation only needs to be done once per mesh and the resulting positions can be saved and loaded for future use. Note that while the implementation can be applied to any mesh, it is optimized for vascular structures and may not produce accurate results for non-vascular meshes.

\textbf{Implementation}. We follow the guidelines by Cech et al. \cite{cech_2020} and employs a specialized C\# matrix math library to handle complex matrix calculations. To reduce computation time, a partial C\# wrapper around the widely used Eigen \cite{eigenweb} library is utilized, utilizing the LDLT solver as suggested by Cech et al. and optimized for a sparse storage format. For endpoint localization, attach both a mesh filter and endpoint script to the desired object. The script will copy the mesh's faces and vertices and successively carry out skeletonization, endpoint localization, and root point identification. This process will run on a background thread as it is not suitable for real-time computation. For better visibility of endpoint locations, an empty GameObject is created for each endpoint to mark its position.

%% file: chapter/11_discussion.tex
\section{Discussion}

We have presented several visualizations to evaluate both egocentric and exocentric distances and provided an implementation of each visualization using Unity. Unity is a widely used game development platform, that has gained popularity in the visualization community due to its modular design. While Unity offers substantial support for standard computer graphics functions, it may restrict users in some cases. Alternatively, utilizing other graphics frameworks like OpenGL offers more flexibility for creating custom visualizations, but with a trade-off of strongly increased programming effort.

\subsection{Expandability and Limitations}
By providing a unified and cross-platform application that contains multiple visualizations for egocentric and exocentric distance estimations, we want to increase accessibility to them and thus increase their potential use in different projects. Unity was chosen for this as it requires minimal setup effort. Other applications may not be cross-platform or may require significantly more time to get started.
While the work is generalizable and theoretically executable in various languages, some effort must be made to convert Unity's built-in variables. Variables such as view direction, vertex position, or model, view, and projection matrices are provided automatically. Although these features are very convenient in most cases, they can be problematic as they cannot be manipulated prior to the shader phase.
A major problem with our implementation was the limitation of point-based rendering. Unity does not support the ability to apply shaders directly to point values; a triangulated mesh is always required. We solve this problem by creating meshes from our point values and then interpreting them as points again in the geometry shader.

The majority of the visualizations can be extended to different meshes and applications, as the associated shaders and scripts can be applied to various objects. However, some visualizations, such as Supporting Lines or Arrow Glyphs, require a list of specific locations to create additional objects. This list can be selected manually or calculated in a pre-processing step, depending on the user's preference. To guide the user, we have provided a pre-calculated point selection and the newly implemented endpoint selection that can be used to position the glyphs even if the mesh is changed.

\subsection{Visualization Selection}
To support the selection of a visualization technique, we included Table \ref{tab:overview}. It provides the occupied information channels, the attentive phase, the distance category, and the system availability. 
We can see which information channels are still available and adjust the visualization or encode additional information. For example, the \textit{heatmap} encodes the exocentric distance using color, while the \textit{isolines} encode the exocentric distance using shape. Since these do not interfere with each other, both can be combined to increase the perception of distance.

In other cases, it is not possible to use the mesh surface to encode information, and we have to switch to glyph-based visualization. If this is also not possible due to spatial constraints, we can move to \textit{illustrative visualizations}. Using this or a similar approach, we can simplify the selection of appropriate visualizations. Providing implementations for each visualization also lowers the threshold to actively use one of the techniques.

%% file: chapter/12_conclusion.tex
\section{Conclusion and Future Work}

In this work, we present a total of sixteen different visualizations and provide detailed information about endpoint detection, which can be used on any vascular model. Glyph-based approaches are often constrained by the selection of glyph locations, which determine their placement. To overcome this limitation, we integrated endpoint detection into our workflow.
This automated approach allows for a more efficient and accurate glyph location selection process. We furthermore discuss fundamental, surface-based, auxiliary, and illustrative visualization techniques, as well as data-driven scalar and vector field visualization. Each technique can help to better understand underlying data and improve estimations of egocentric and exocentric distance information. We also include an overview of occupied visual cues for each visualization, which can promote different combinations or extensions of each technique. Our goal is to continue updating this framework to promote further research in vascular visualization and psycho-physical studies. Overall, this work provides a comprehensive overview and implementation to encourage continued research in this field.